\documentclass{aa}  

\usepackage{txfonts}
\usepackage{graphicx}	% Including figure files
\usepackage{amsmath}	% Advanced maths commands
\usepackage{amssymb}	% Extra maths symbols
\usepackage{comment}
\usepackage{soul}
\usepackage{color}
\usepackage{xcolor}
\usepackage{hyperref}
\usepackage{multirow}
\def\lsun{{\rm L_{\odot}}}
\def\msun{{\rm M_{\odot}}}
\def\rsun{{\rm R_{\odot}}}

\def\msunyr{{\rm M_{\odot}~\rm{yr}^{-1}}}

\defcitealias{2020OlivaKuiper}{OK20}

\begin{document}

   \title{From Fragments to Flares: Migration, Tidal Disruption, and Observable Bursts in Massive Protostellar Disks}

   % \subtitle{I. Overviewing the $\kappa$-mechanism}

   \author{ Vardan Elbakyan\inst{1},
            Rolf Kuiper\inst{1},
           André Oliva\inst{2}, 
            Verena Wolf\inst{3},
            Jochen Eislöffel\inst{3},
            Bringfried Stecklum\inst{3},
            Christian Andreas\inst{3}
          }
   \institute{Fakultät für Physik, Universität Duisburg-Essen, Lotharstraße 1, 47057 Duisburg, Germany; 
   \email{vardan.elbakyan@uni-due.de} 
   % \and
   %   Research Institute of Physics, Southern Federal University, Rostov-on-Don 344090, Russia   
    \and
     Space Research Center (CINESPA), School of Physics, University of Costa Rica, San José, Costa Rica;  
    \and  
    Thüringer Landessternwarte Tautenburg, Sternwarte 5, 07778 Tautenburg, Germany
}

    \titlerunning{Fragment-driven bursts}
    \authorrunning{Elbakyan et al.}

   \date{Received January 14, 2026; accepted TBD}

% \abstract{}{}{}{}{} 
% 5 {} token are mandatory
 
  \abstract
   {Gravitational fragmentation in massive protostellar disks can lead to the formation of bound gaseous clumps whose inward migration and disruption may trigger strong accretion outbursts. Yet, the role of numerical resolution and inner boundary treatment in shaping the burst properties remains poorly understood.}
   {We investigate how resolving the inner few astronomical units of a massive protostellar disk affects the migration, disruption, and accretion signatures of an inward-moving fragment. In particular, we aim to determine whether the predicted burst strength and duration depend on the adopted sink cell size.}
   {We present a new three-dimensional radiation-hydrodynamic simulation of a $\sim$5$\msun$ protostar surrounded by a self-gravitating disk, comparing the original 30~AU sink model to a refined model with a 1~AU sink that resolves the inner disk. The resulting gas structures are post-processed with radiative transfer calculations to derive synthetic photometry and multi-band images.}
   {Both simulations produce a major accretion burst as a migrating fragment is tidally disrupted, but their detailed behavior differs markedly. The refined model shows faster migration, a complete tidal disruption of the fragment, and a shorter, sharper outburst (more consistent with observations) with nearly the same peak accretion rate as the 30~AU model, which yields a broader, smoother event. The refined run produces much stronger near- and mid-infrared emission, reflecting the formation of a compact, hot inner disk.}
   {Resolving the inner few AU qualitatively changes the dynamics and observable appearance of fragment-driven bursts. Diffuse fragment disruption can reproduce decade-long events, but the much shorter ($<$3~yr) bursts observed in some massive protostars likely require the tidal disruption of more compact objects such as second Larson cores. Our trajectory analysis indicates that second Larson cores can migrate sufficiently close to the star to be tidally destroyed, offering a plausible mechanism for the fastest FU-Ori–like bursts observed in massive protostars.}


   \maketitle
%
%-------------------------------------------------------------------

%%%%%%%%%%%%%%%%%%%%%%%%%%%%%%%%%%%%%%%%%%%%%%%%%%%%%%%%%%%%%%%%%%%%%%%%%%%%%%
%%%%%%%%%%%%%%%%%%%%%%%%%%%%%%%%%%%%%%%%%%%%%%%%%%%%%%%%%%%%%%%%%%%%%%%%%%%%%%
%%%%%%%%%%%%%%%%%%%%%%%%%%%%%%%%%%%%%%%%%%%%%%%%%%%%%%%%%%%%%%%%%%%%%%%%%%%%%%

\section{Introduction}

Disk fragmentation is a fundamental process in star formation, occurring in disks around protostars of any masses \citep{2025BeutherKuiper}.
Theoretical models predict that high-mass protostellar disks, due to their high gas densities and rapid accretion rates, are particularly prone to gravitational instabilities, leading to the formation of self-gravitating fragments \citep{2006KratterMatzner, 2010Kratter, 2011Kuiper, 2016Kratter}. In recent years, a growing number of observations have confirmed the presence of disk-like structures around intermediate- to high-mass protostars \citep{2010Kraus, 2016Chen, 2017Cesaroni, 2016Ilee, 2018Ilee, 2018Beuther, 2018Maud, 2019Maud, 2019Zapata, 2019Sanna, 2021Sanna, 2025Sanna, 2015Johnston, 2020Johnston, 2020Guzman, 2019Moscadelli, 2021Moscadelli, 2021Suri, 2019Ahmadi, 2023Ahmadi, 2023Burns, 2025Bayandina}, supporting the idea that disk-mediated accretion is a standard mechanism for high-mass star formation. Furthermore, recent studies have resolved arc-, ring-, or spiral-like substructures in these disks, likely formed due to gravitational instabilities \citep{2019Maud, 2019Zapata, 2020Johnston, 2021Sanna, 2023Burns}.  

One of the most important consequences of disk fragmentation is its role in episodic accretion and luminosity bursts. In low-mass protostars, the episodic accretion of disk fragments has been proposed as a key mechanism behind FU Ori-type outbursts \citep{2010ForganRice, 2012NayakshinLodato, 2010VorobyovBasu, 2015VorobyovBasu, 2018VorobyovElbakyan, 2019VorobyovElbakyan, 2018ZhaoCaselli}. Similar accretion bursts have recently been observed in high-mass protostars \citep{2015Tapia, 2017Caratti, 2017Hunter, 2019Proven-Adzri, 2021Hunter, 2021Stecklum, 2021Chen, 2022Zhang, 2023Fedriani, 2024WolfStecklum, 2025ContrerasPena} (many of which are considerably shorter than classical FU Ori–type events), suggesting that fragment-driven accretion bursts may be a universal feature of star formation across different mass regimes \citep{2025BeutherKuiper}. Understanding the evolution of disk fragments, their interaction with the central star, and their contribution to accretion bursts is therefore essential for developing a complete picture of protostellar evolution. 

Massive stars are often found in binary or multiple systems \citep[e.g.,][]{2012Sana}. 
While both core fragmentation (during the initial collapse of the natal cloud) \citep{2024Li} and disk fragmentation (during later dynamical evolution) \citep{2025Li} may contribute to multiplicity, the latter is increasingly favored as the primary mechanism for forming close companions.
Recent studies suggest that binaries may initially form at large separations and later migrate inward due to interactions with a remnant accretion disk or other young stellar objects in the system \citep[e.g.,][]{2024RamirezTannus}. 
A critical challenge lies in explaining how fragments survive migration to form compact, stable systems. Fragments approaching the central star face tidal disruption, especially if they remain extended gas clumps.

One possible pathway for fragment survival is the formation of a second, more compact core at its center.  As the fragment contracts and its central temperature rises above the $\mathrm{H}_2$ dissociation threshold $\sim$2000~K, molecular hydrogen dissociation can trigger a second collapse and the birth of a dense, hydrostatic “second” Larson core \citep[e.g.,][]{1969Larson, 2000Masunaga, 2013Tomida, 2020Bhandare}. Recent 1D–2D collapse calculations quantify the mass–radius relation and thermal evolution of these objects over a wide range of initial cloud masses \citep[e.g.,][]{2013Vaytet, 2017Vaytet, 2018Bhandare, 2020Bhandare}. A compact second core, with characteristic sizes much smaller than the parent first core (from a few $\rsun$ up to $\sim$sub-AU scales in many models), is far more resistant to tidal shear than the diffuse fragment envelope and therefore can survive to much smaller orbital radii before disruption.  Once formed, such second cores may evolve into stellar companions (via continued accretion and contraction) or into substellar objects depending on their accretion history and migration \citep[e.g.,][]{2016Kratter,2020OlivaKuiper}; conversely, the tidal–downsizing scenario shows how migrating, partially collapsed fragments can produce lower-mass (planetary / brown-dwarf) outcomes if disrupted earlier \citep[e.g.,][]{2010NayakshinB, 2015Nayakshin}. The precise conditions that determine whether a second core becomes a close stellar companion, a brown dwarf, or is tidally destroyed remain active areas of research.

A key limitation in previous numerical studies of disk fragmentation is the treatment of the inner disk. As shown in \citet{2023ElbakyanNayakshin}, insufficient resolution in the inner tens of astronomical units can lead to overestimated accretion rates onto the star, artificially amplifying accretion bursts when fragments cross the sink cell boundary. Related studies have similarly highlighted how the choice of inner boundary/sink radius and the absence of a resolved inner disk can bias accretion variability and feedback \citep[e.g.,][]{2019MeyerVorobyovElbakyan, 2009Bate}. A complementary code-comparison by \citet{2023Mignon-Risse} shows that, although RAMSES and PLUTO agree on global collapse and fragmentation, innermost-disk outcomes (fragment number, dynamics, and multiplicity) are sensitive to numerical choices (grid geometry and sink prescriptions) and that running RAMSES with a single central sink brings its results closer to PLUTO, implying that inner-boundary and accretion subgrid models, rather than the code itself, govern inner-disk behavior. In addition, thin–disk calculations demonstrate that the mass transport rate through the innermost region strongly controls the global disk evolution, fragmentation propensity, dust growth and burst behavior, underlining that the inner boundary model is not a neutral numerical choice but a physically important ingredient \citep[e.g.,][]{2019VorobyovSkliarevskii}. In reality, a smaller sink cell allows for a more accurate treatment of fragment migration, interaction with the inner disk, and potential survival. Resolving the inner disk is therefore essential for correctly modeling accretion variability, disk fragmentation, and the formation of secondary objects.  

In this study, we investigate the passage of a disk fragment through a significantly smaller sink cell (1 AU) compared to previous models that used a 30 AU sink cell \citep[][hereafter OK20]{2020OlivaKuiper}. This allows us to track the evolution of the fragment as it approaches the central star, examining whether it forms a compact second core,  survives migration, or undergoes tidal disruption. 
We quantify the impact of these outcomes on accretion variability and observable signatures by computing synthetic continuum images and light curves, and we assess the consequences for close binary and multiple–system formation.
By resolving the inner disk at sub-AU scales, we aim to provide new insights into the final stages of disk fragmentation and its role in shaping the formation of massive stars, as well as the observational manifestations of these processes.

%%%%%%%%%%%%%%%%%%%%%%%%%%%%%%%%%%%%%%%%%%%%%%%%%%%%%%%%%%%%%%%%%%%%%%%%%%%%%%
%%%%%%%%%%%%%%%%%%%%%%%%%%%%%%%%%%%%%%%%%%%%%%%%%%%%%%%%%%%%%%%%%%%%%%%%%%%%%%
%%%%%%%%%%%%%%%%%%%%%%%%%%%%%%%%%%%%%%%%%%%%%%%%%%%%%%%%%%%%%%%%%%%%%%%%%%%%%%
\section{Numerical model}\label{sec:model}
\subsection{Physics considered}

Our calculations solve three-dimensional hydrodynamics with self-gravity and radiative transfer using the PLUTO code \citep{2007Mignone}. The gas is treated as an ideal fluid, and the model includes a two-temperature flux-limited diffusion treatment of thermal radiation together with a frequency-dependent stellar irradiation term. The self-gravity of gas is computed throughout the domain. For brevity, we do not reproduce the full set of equations and numerical implementation here. The original formulation is described in detail in \citetalias{2020OlivaKuiper}, and a compact summary of the specific settings used in this work is given in Appendix~\ref{sec:model}. Sections~\ref{sec:grid} and \ref{sec:init_bound_cond} summarize the grid, restart, and boundary choices that differ from the OK20 setup.

%%%%%%%%%%%%%%%%%%%%%%%%%%%%%%%%%%%%%%%%%%%%%%%%%%%%%%%%%%%%%%%%%%%%%%%%%%%%%%

\subsection{Extended computational grid and inner boundary} \label{sec:grid}
The domain uses spherical coordinates ($\theta=0\ldots\pi$, $\phi=0\ldots2\pi$, $r_{\rm max}=0.1\ \mathrm{pc}$) with logarithmic radial spacing to concentrate resolution near the star, cosine-weighted polar spacing to enhance the midplane, and uniform azimuthal spacing. Starting from the grid of model \textit{x8} in \citetalias{2020OlivaKuiper}, we shift the inner boundary from 30~AU to 1~AU (increasing the number of radial cells by 140) to resolve the inner few AU and retain smooth logarithmic spacing. The final mesh is $408\times81\times256$ (radial, polar, azimuthal), providing sub-AU cells inside $\sim$40~AU and substantially improving capture of inner-disk dynamics while keeping the outer domain consistent with the original run.
Near $R\simeq 1$~AU the grid spacing is $\Delta r \simeq 2.4\times10^{-2}$~AU, and the midplane-enhanced polar spacing corresponds to $\Delta \theta \simeq 2.5\times10^{-2}$~rad, i.e. $\Delta z \approx R\,\Delta\theta \simeq 2.5\times10^{-2}$~AU at 1~AU (about 40 cells per AU in both $r$ and $z$ near the midplane). For the disk aspect ratios measured in our model, $h\equiv H/R \simeq 0.05$--0.13, this corresponds to $\sim$2--5 cells per pressure scale height in the polar direction at $R\simeq 1$~AU. This vertical resolution is adequate for our analysis of the global disk dynamics and midplane evolution, although it is not intended as a high-precision study of the detailed vertical hydrostatic structure at 1~AU.

%%%%%%%%%%%%%%%%%%%%%%%%%%%%%%%%%%%%%%%%%%%%%%%%%%%%%%%%%%%%%%%%%%%%%%%%%%%%%%
\subsection{Initial and Boundary Conditions}  \label{sec:init_bound_cond}
The initial conditions in our simulations are based on the setup described in \citetalias{2020OlivaKuiper}; however, instead of starting from the collapse of a spherically symmetric molecular cloud, we initialize the system using the physical properties of the disk at $t_0=8580$~years after the onset of the molecular cloud collapse in the original model, which is $\sim$40~yr before the fragment’s passage through the original 30~AU sink. 
In that model, once a fragment entered the sink cell, its mass was counted as accreted onto the central massive protostar, and its subsequent evolution could not be followed. This passage triggered a strong accretion outburst, accompanied by the destruction of the fragment (see Sect.~\ref{sec:outburst}).

With the inner boundary moved to $r=1$~AU, we can follow the fragment’s close passage, its interaction with the inner disk, and its tidal destruction in much greater detail. This refined setup captures key processes that the larger sink missed and thus permits a more realistic study of accretion bursts, inner-disk instabilities, and the conditions that favor the formation of close companions. The main reason this was not feasible in OK20 is computational cost: reducing the inner boundary from 30 to 1 AU shortens the timestep by more than two orders of magnitude, making such simulations vastly more expensive.

To accommodate the extended grid, 
we extrapolate all relevant physical properties (e.g., density, temperature, and velocity) into these newly introduced cells.
The extrapolation procedure and the resulting initial density maps are described in Appendix~\ref{sec:extrapol_appendix}.
In what follows, we refer to the original 30 AU–sink simulation of \citetalias{2020OlivaKuiper} as the “OK20 model”, and to the new run with a 1 AU sink cell as the “refined model”.

The boundary conditions remain consistent with those in \citetalias{2020OlivaKuiper}, with semi-permeable inner and outer boundaries that allow material to flow out of the computational domain but prevent inflow. The mass within the sink cell is considered accreted onto the central star, influencing its growth and luminosity evolution, and no explicit outflow or jet feedback from the accreting material is included.

%%%%%%%%%%%%%%%%%%%%%%%%%%%%%%%%%%%%%%%%%%%%%%%%%%%%%%%%%%%%%%%%%%%%%%%%%%
%%%%%%%%%%%%%%%%%%%%%%%%%%%%%%%%%%%%%%%%%%%%%%%%%%%%%%%%%%%%%%%%%%%%%%%%%%

\section{Fragment migration and disruption}

\subsection{Fragment evolution in the OK20 model}

Disk fragmentation around massive protostars produces dense fragments that may migrate inward and be tidally disrupted.  In this section, we follow the fate of one such fragment from its birth in the outer spiral arm through its inward journey and final disruption near the central sink cell. 

\begin{figure}
    \centering
    \includegraphics[width=1\linewidth]{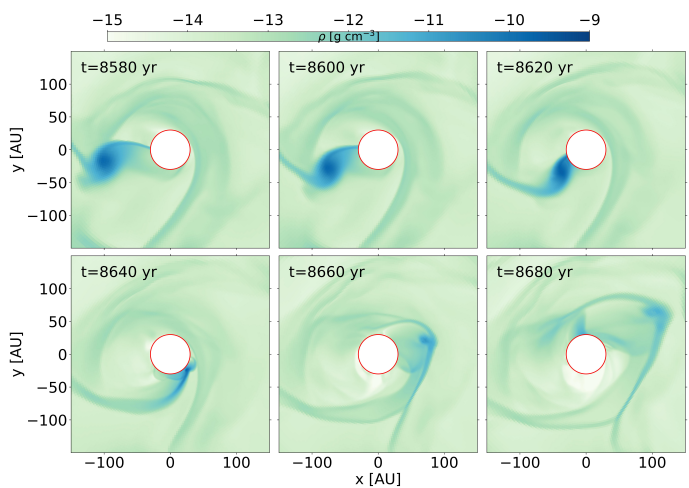}
    \caption{Midplane gas density maps illustrating the tidal disruption of the fragment at six selected times in the \citetalias{2020OlivaKuiper} simulation. The red circle denotes the 30~AU sink cell radius.}
    \label{fig:rho_6plot_orig}
\end{figure}

Figure~\ref{fig:rho_6plot_orig} presents six midplane density maps from the OK20 model as the fragment passes by the 30 AU sink. The majority of its mass is accreted in the first passage by the sink cell, yet a trailing spiral arm survives, stretching toward the sink and forming an elongated stream, which later accretes onto the central star.

\begin{figure}
    \centering
    \includegraphics[width=1\linewidth]{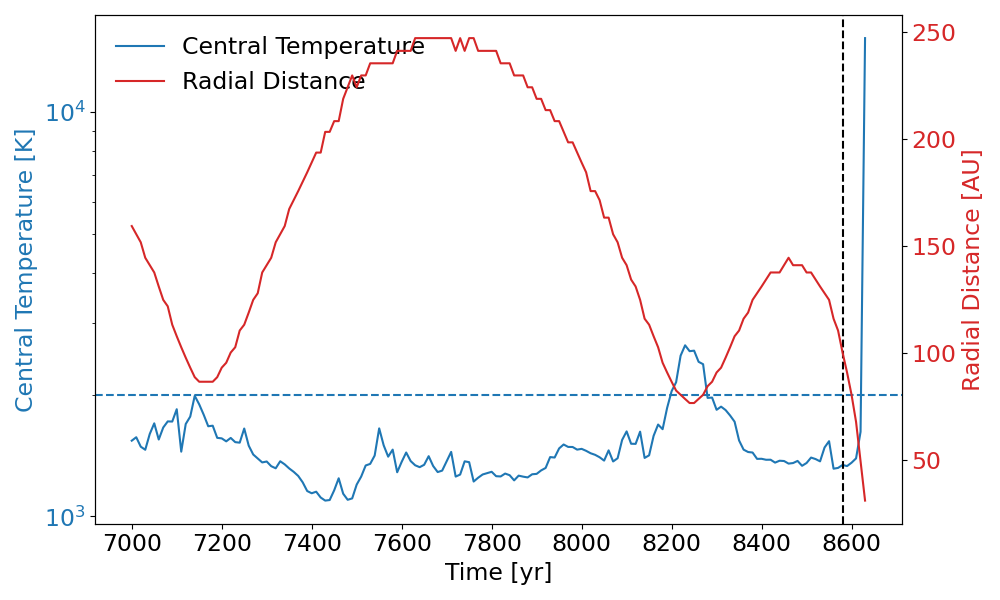}
    \caption{Evolution of the fragment’s central temperature $T\rm{_c}$ (blue, left axis) and radial distance from the star $r\rm{_c}$ (red, right axis) as a function of time. The dashed blue line marks the critical temperature for H$_2$ dissociation ($\sim$2000~K), while the vertical dashed line indicates the initial time instance for the refined model.}
    \label{fig:Tpeak_radius_orig}
\end{figure}

Building on the morphological evolution shown in Figure~\ref{fig:rho_6plot_orig}, we now turn to its orbital and thermal evolution. Using our fragment‐tracking algorithm (see Appendix~\ref{sec:fragment_contour}), we follow the fragment’s radial distance and the central temperature. Figure~\ref{fig:Tpeak_radius_orig} shows the fragment’s radial distance from the star (red curve, right axis) alongside its central temperature (blue curve, left axis) from 7000~yr, through the restart at 8580~yr (vertical dashed line), and continuing until the fragment passes the sink cell and is disrupted/accreted. As the fragment’s periastron decays inward, compressional heating due to gravitational collapse drives its core temperature upward. At $\sim8200$~years, the central temperature crosses the H$_2$ dissociation threshold of $\approx$2000~K (horizontal dashed line). This thermodynamic shift heralds the formation of a bona fide secondary object, a hydrostatically supported second Larson core \citep{2000MasunagaInutsuka, 2020Bhandare}, that will continue its own inward migration under disk torques. We analyze the dynamics and survivability of this nascent core in Sect.~\ref{sec:second_core}.

\begin{figure}
    \centering
    \includegraphics[width=1\linewidth]{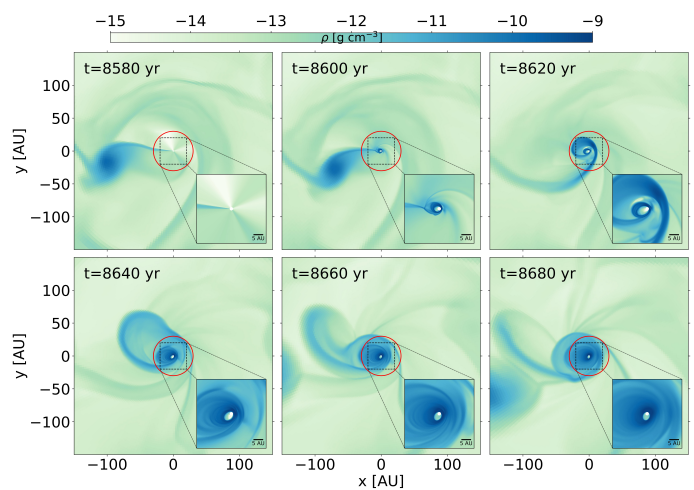}
    \caption{Midplane gas density during the tidal disruption of the fragment, shown at six different times. The red circle indicates the 30~AU sink cell radius used in the \citetalias{2020OlivaKuiper}. Insets in the lower right of each panel provide a zoomed view of the inner 40~AU~$\times$~40~AU region, with a 5~AU scale bar for reference. Dashed rectangles on the main panels denote the inset regions.}
    \label{fig:rho_6plot_corr}
\end{figure}

%%%%%%%%%%%%%%%%%%%%%%%%%%%%%%%%%%%%%%%%%%%%%%%%%%%%%%%%%%%%%%%%%%%%%%%%%%%%%%%%%%%%%%%%%%%%%%%%%

\subsection{Fragment evolution in the refined model}

Figure~\ref{fig:rho_6plot_corr} presents six snapshots at the same evolutionary times as in Figure~\ref{fig:rho_6plot_orig}, now with a 1~AU sink cell. Shortly after the restart of the simulation with the extended grid, the density structure inside the extrapolated region relaxes and comes into equilibrium with the original disk beyond 30~AU. As the fragment continues its inward migration in the refined run, it steadily sheds mass along a pronounced inner spiral arm that channels gas from its outer envelope into the newly resolved few‐AU region. This accretion stream first winds into a compact, eccentric ring at radii of a few AU, where it feeds the 1~AU sink cell. As the fragment approaches periastron, tidal forces overwhelm its self‐gravity, tidally disrupt it, and tear the remaining bound material into an elongated filament. This filament initially expands outward under the same slingshot action that disrupted it, then curves back inward in a broad spiral. Over the following orbits, the returning debris settles into a dense inner disk surrounding the sink cell. This newly formed inner disk is itself eccentric, with $e\sim0.65$. Because the elongated filament and the dense disk lie almost entirely within 30~AU, treated as a single unresolved sink in the OK20 model, its detailed morphology and the timing of these feeding episodes were inaccessible without our extended‐grid treatment.

\begin{figure}
    \centering
    \includegraphics[width=1\linewidth]{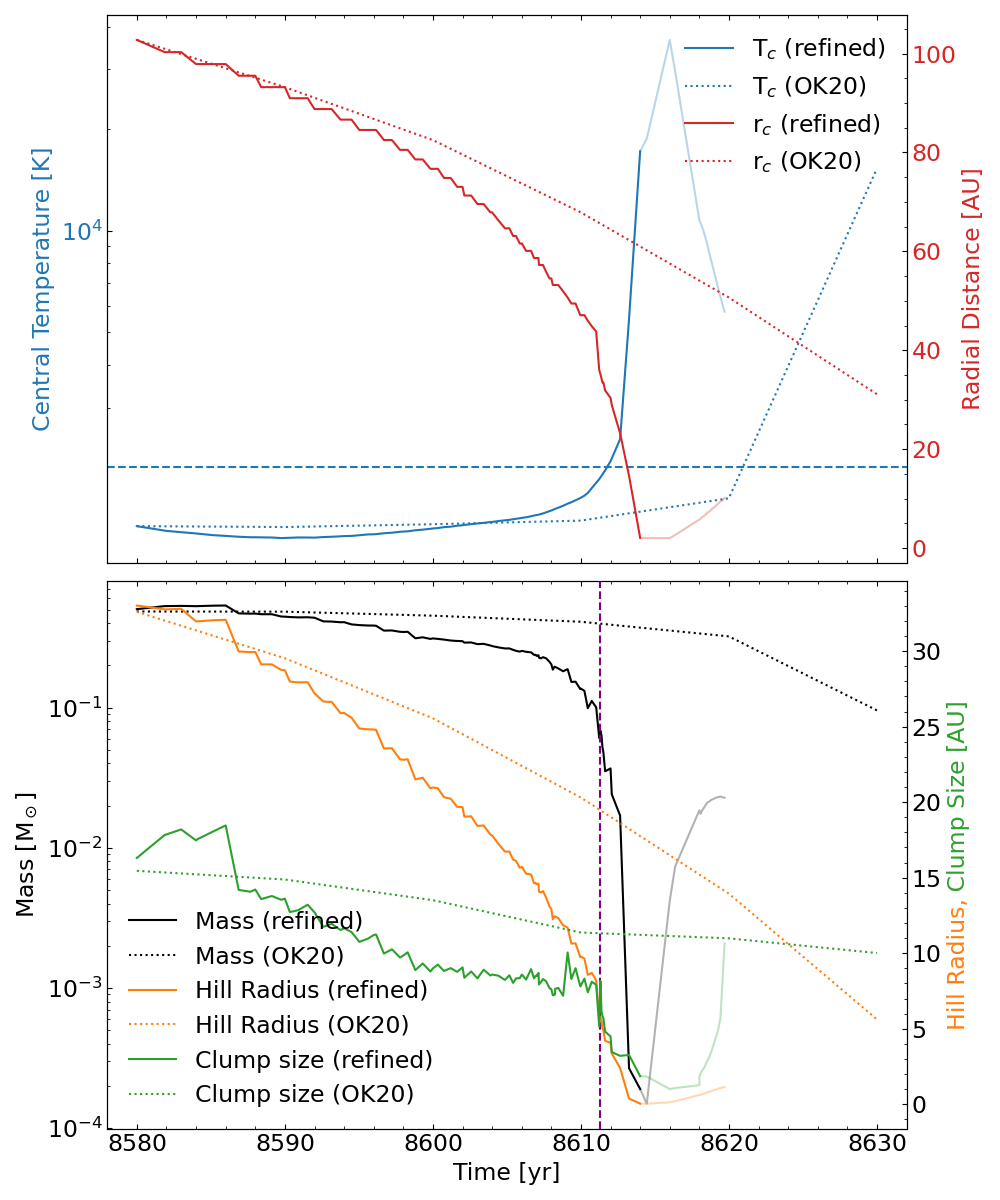}
    \caption{Time evolution of fragment properties for the refined (solid; semi-transparent after disruption) and OK20 (dotted) tracks. \textbf{Top:} central temperature (left axis) and radial distance from the star (right axis); the horizontal dashed line marks $T = 2000$~K. \textbf{Bottom:} fragment mass  (left axis), Hill radius, and mean fragment radius (right axis); the vertical dashed line marks when the mean fragment radius equals the Hill radius.}
    \label{fig:Tpeak_radius}
\end{figure}

To quantify how the fragment evolves as it spirals inward and is tidally disrupted, we plot in Figure~\ref{fig:Tpeak_radius} its orbital radius (red), central temperature (blue), bound mass (black), Hill radius (orange), and effective size (green) for both the refined and OK20 models. 

In the refined run (solid lines in Figure~\ref{fig:Tpeak_radius}; semi-transparent after disruption), the fragment migrates from $\sim$100~AU down to a few AU (effectively reaching the sink cell — note the flattened minimum plateau of the red curve). Compressional heating due to gravitational collapse drives the core temperature from $\sim$1500~K up to a few $\times10^4$~K immediately prior to disruption, producing the sharp blue spike near $\sim$8612~yr. At that point the fragment is violently stripped: the bound mass drops by more than two orders of magnitude, the Hill radius shrinks from $\sim$10~AU to $\lesssim$1~AU, and the measured physical radius falls to a few AU. A small, denser remnant survives the encounter and partially re-accretes, producing the gradual rebound seen in mass, size and Hill radius after $\sim$8615~yr.

By contrast, the OK20 run (dotted lines) exhibits a much smoother evolution. The fragment steadily migrates from $\sim$100~AU into the 30 AU sink; its Hill radius contracts monotonically from $\sim$33~AU to $\sim$5~AU, and its physical radius decreases only slowly (from $\sim$15 to $\sim$10~AU). The central temperature in OK20 does rise — it crosses the H$_2$ dissociation threshold ($\sim$2000~K) at $\sim$50~AU (vs $\sim$30~AU in the refined run) and reaches a few $\times10^4$~K at accretion — but the increase is less abrupt than in the refined calculation. Mass loss is comparatively gradual: the fragment’s mass declines from $\sim$0.5$\msun$ to $\sim$0.08$\msun$ by the time it is swallowed by the 30~AU sink, rather than the near-complete depletion (to a few $10^{-4}$$\msun$) recorded in the refined run.

Overall, both models show that the infalling fragment undergoes intense compressional heating due to gravitational collapse as it approaches the star, with the central temperature rising steeply to above $10^4$~K just before disruption or accretion. However, the subsequent evolution differs fundamentally. In the refined model, most of the fragment’s mass is stripped and dispersed within the inner disk, leaving only a tiny remnant that partially re-accretes. In the OK20 model, by contrast, the fragment remains largely intact as it crosses the 30~AU sink, losing only a fraction of its initial mass. 
Thus, despite comparable heating histories, the two models channel the fragment’s mass very differently: the refined run disperses it into the inner disk, while the OK20 run delivers it intact to the star through the large sink.

This distinction has important implications for how accretion bursts manifest. Because roughly the same total mass is transported inward in both models, the stellar accretion rates and corresponding burst amplitudes may appear similar when measured at the sink boundary (see Sect.~\ref{sec:outburst}). Importantly, this implies that a large (30 AU) sink captures the global mass budget and the bulk star–disk–envelope mass exchange reasonably well, so OK20 remains useful for studies of overall accretion and large-scale feedback. However, the physical and observational consequences differ markedly. In the refined model, the mass is first deposited into the inner disk, where it can be reprocessed and radiated, leading to an outburst and a luminous inner disk that persists after the disruption. In the OK20 case, that same material is accreted instantaneously by the star, producing a burst without the formation of a sustained hot inner structure.

These differences show that resolving the inner few AU is essential, because with a smaller sink the diffuse fragment can migrate closer to the star and produce qualitatively different mass deposition and radiative signatures (see Sect.~\ref{sec:radmc}) than when the same mass is swallowed by a large, unresolved sink.  We therefore emphasise that accurate prediction of burst morphology and observables requires resolving the inner disk region at sub-AU scales.

%%%%%%%%%%%%%%%%%%%%%%%%%%%%%%%%%%%%%%%%%%%%%%%%%%%%%%%%%%%%%%%%%%%%%%%%%%

\subsection{Accretion Burst Properties} \label{sec:outburst}

\begin{figure}
    \centering
    \includegraphics[width=1\linewidth]{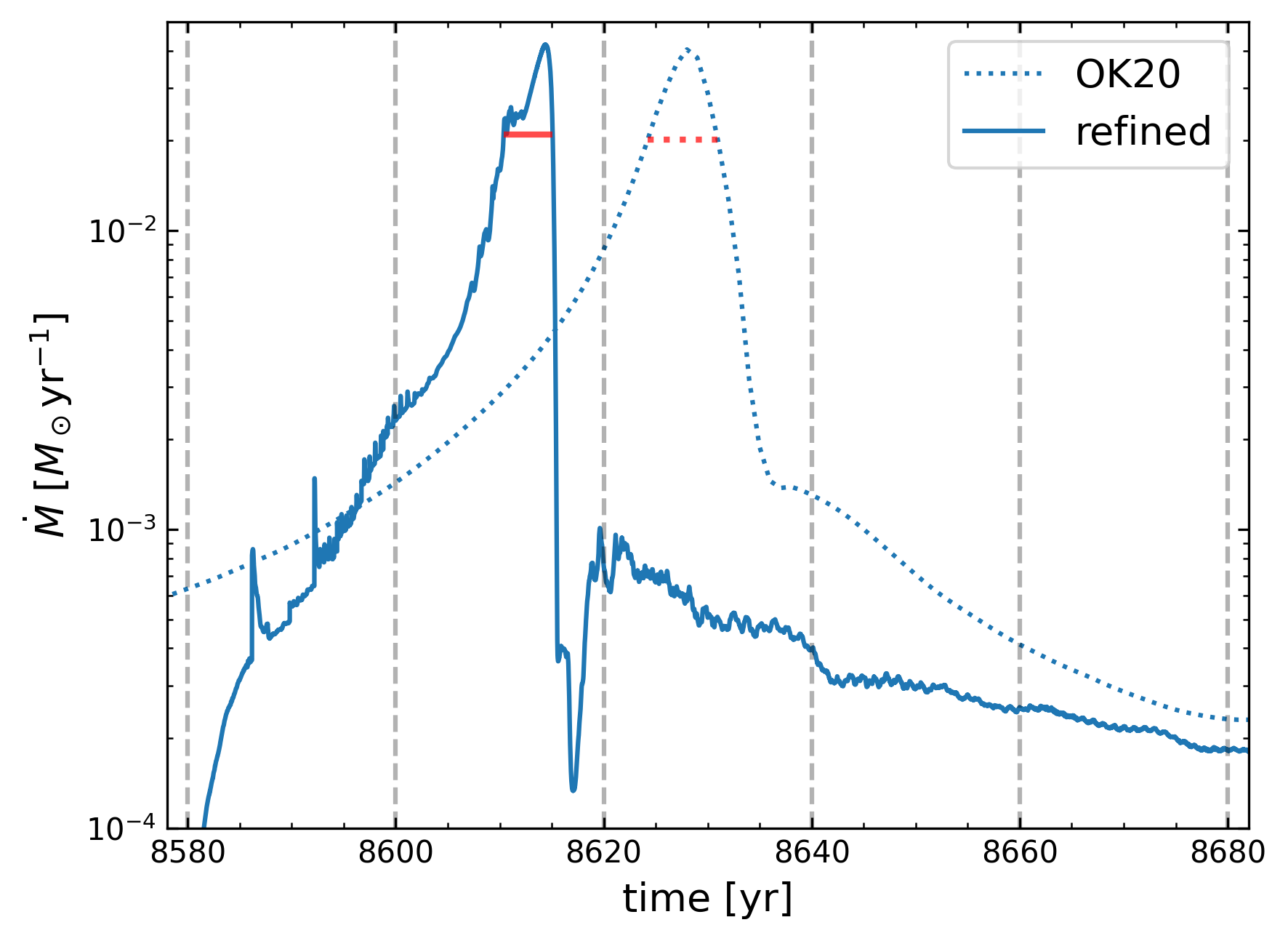}
    \caption{Accretion rate onto the central star as a function of time for two models: OK20 (dotted line) and refined (solid line). The vertical dashed lines mark the time instances shown in Figure~\ref{fig:rho_6plot_corr}. For each model, the full width at half maximum (FWHM) duration of the burst are indicated by horizontal red lines.}
    \label{fig:mdot}
\end{figure}

We now examine how the fragment’s tidal disruption translates into a mass accretion burst onto the central protostar, by comparing the accretion histories of the original OK20 model and our refined, high‐resolution run.  
In Fig.~\ref{fig:mdot} we show the accretion rate histories for both models. Both the OK20 (dotted line) and refined (solid line) models show a gradual rise in $\dot M$ as the disrupted material streams inward, but the refined run exhibits a noticeably steeper ascent and a much sharper post-peak decline. Immediately after the restart the refined model shows a lower accretion rate because the newly introduced 1~AU sink initially contains little mass and the inner flow must re-establish; the inner region relaxes into a new quasi-steady configuration and accretion through the small boundary ramps up as material is funneled in.

In the refined model, the accretion rate increases steadily over $\sim20$~yr — roughly twice as fast as in the OK20 case — reflecting more efficient delivery of dense material via the inner spiral arm that survives tidal disruption. The spiral arm continuously feeds the inner few-AU region, smoothing out what would otherwise be an abrupt rise. Once the arm and the fragment have fully passed the sink boundary, the supply is suddenly cut off, producing a sharp drop from $\sim4\times10^{-2}$ down to $\sim4\times10^{-4}\,\msunyr$ in about one year. This sharp decline correlates with the near-complete loss of bound mass seen in Figure~\ref{fig:Tpeak_radius}. Meanwhile, loosely bound debris are slung outward by the same tidal forces that disrupted the fragment. These leftovers then fall back in on slightly longer trajectories, producing a secondary bump in $\dot M$ up to $\sim10^{-3}\,\msunyr$ before the supply finally dwindles and large-scale advection and numerical diffusion govern the long‐term decline.

By contrast, the OK20 model exhibits both a slower rise and a more gradual fall. With a 30~AU sink cell, the infalling stream is intercepted farther out and redistributed through a larger unresolved volume.  Consequently, $\dot M$ climbs more gently, peaks at nearly the same level as in refined model, then declines over $\sim8$~yr to $\sim10^{-3}\,\msunyr$ before transitioning into a long, gentle decay. The absence of a sharp post-peak shut-off in OK20 likely reflects both that the fragment retains much of its mass as it crosses the large sink (so there is no abrupt loss of supply) and the averaging and numerical diffusion introduced by the large sink.
Also, the coarser temporal sampling of 1~year (vs. $10^{-3}$~year in refined model) further smooths the curve.

An accretion burst can be characterized by its peak accretion rate $\dot M_{\rm peak}$ and its duration $\Delta t_{\rm burst}$. A common criterion for $\Delta t_{\rm burst}$ is the full‐width at half‐maximum (FWHM), defined as the time interval during which $\dot M$ remains above half of $\dot M_{\rm peak}$. In the OK20 model, the burst reaches a maximum of $4\times10^{-2}\,\msunyr$ and maintains at least half that rate for approximately 6.7~years.  In the refined model, the peak is slightly higher, $4.2\times10^{-2}\,\msunyr$, but the FWHM is shorter, about 4.7~years. Thus, while both simulations channel a similar total mass into the inner region over the entire event (including subsequent accretion of formed inner disk), they differ in where and when that mass is released onto the star: OK20 deposits material directly into the large sink, producing a smoother, more extended sink-accretion episode, whereas the refined run deposits mass into a compact inner disk that is accreted onto the star over a longer inner-disk accretion timescale — yielding a shorter immediate FWHM at the sink but a much longer-lived tail as the inner disk drains.

%%%%%%%%%%%%%%%%%%%%%%%%%%%%%%%%%%%%%%%%%%%%%%%%%%%%%%%%%%%%%%%%%%%%%%%%%%
%%%%%%%%%%%%%%%%%%%%%%%%%%%%%%%%%%%%%%%%%%%%%%%%%%%%%%%%%%%%%%%%%%%%%%%%%%

\section{Impact on the thermal structure of the disk}\label{sec:therma_structure}

\begin{figure}
    \centering
    \includegraphics[width=1\linewidth]{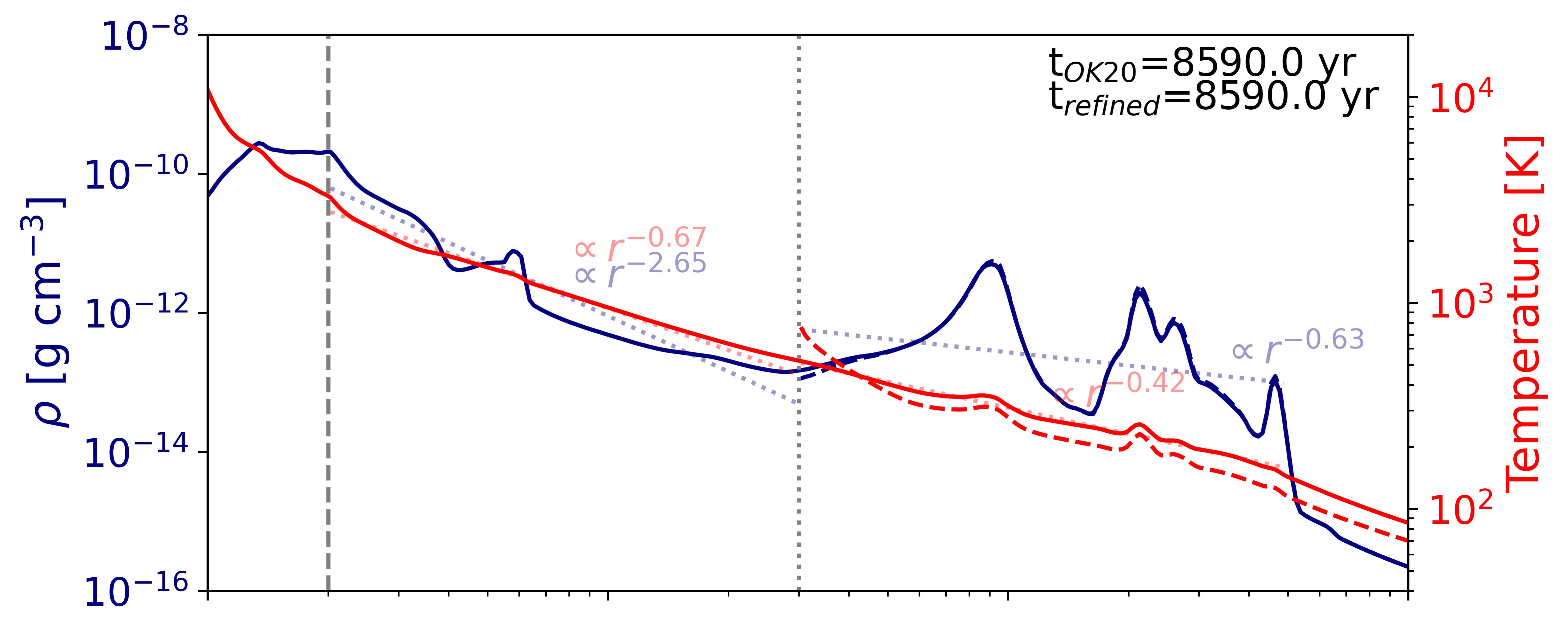}
    \includegraphics[width=1\linewidth]{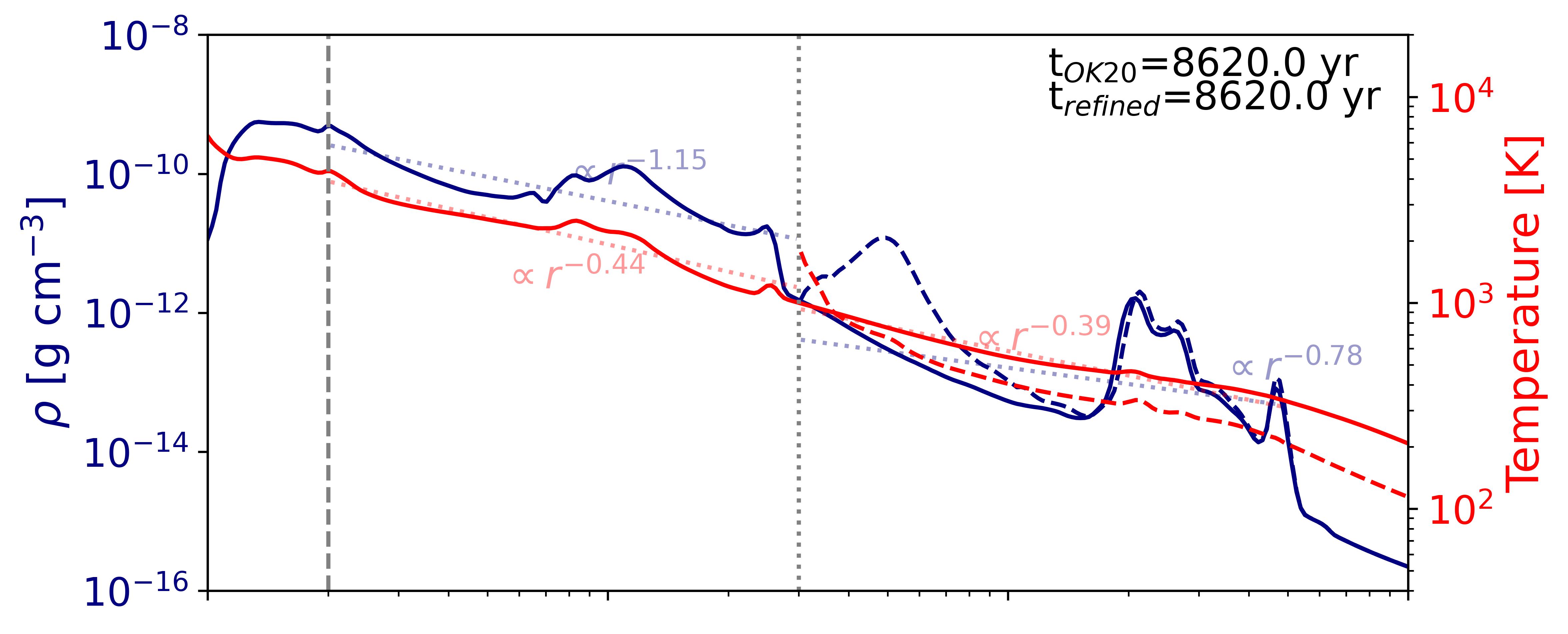}
    \includegraphics[width=1\linewidth]{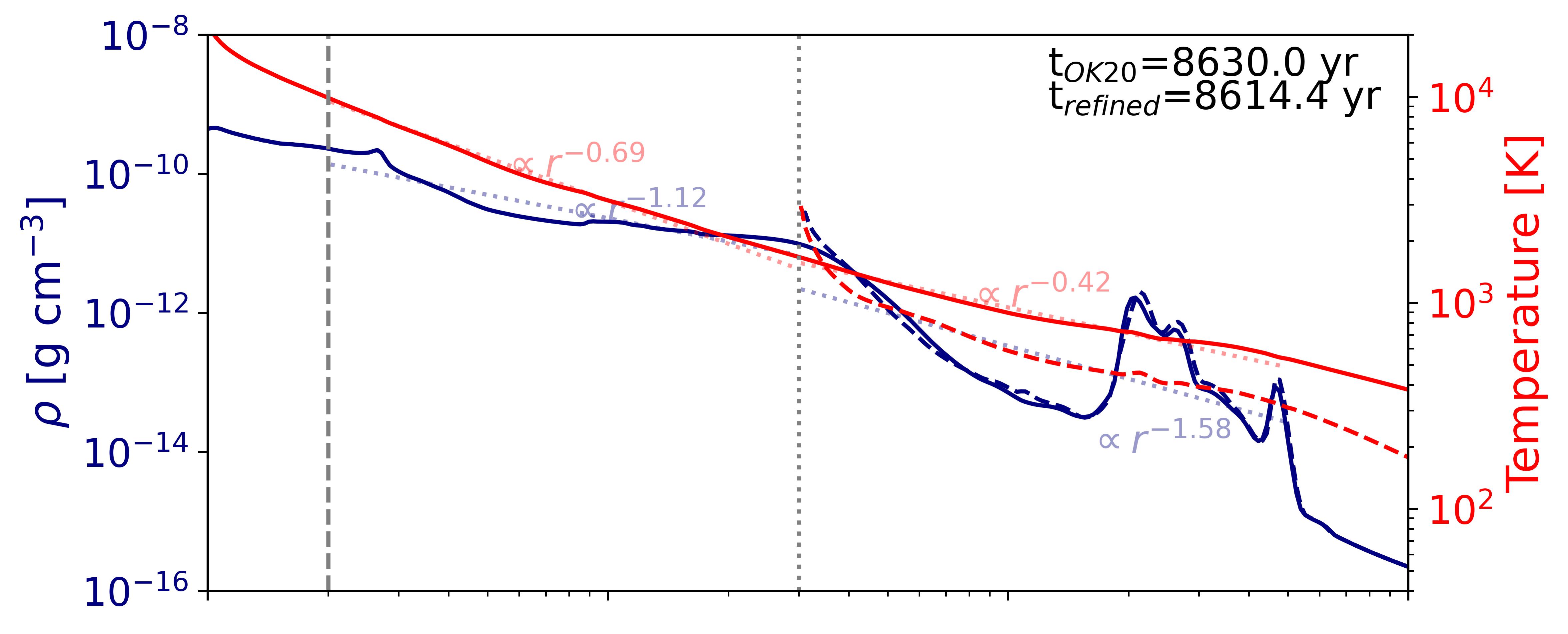}
    \includegraphics[width=1\linewidth]{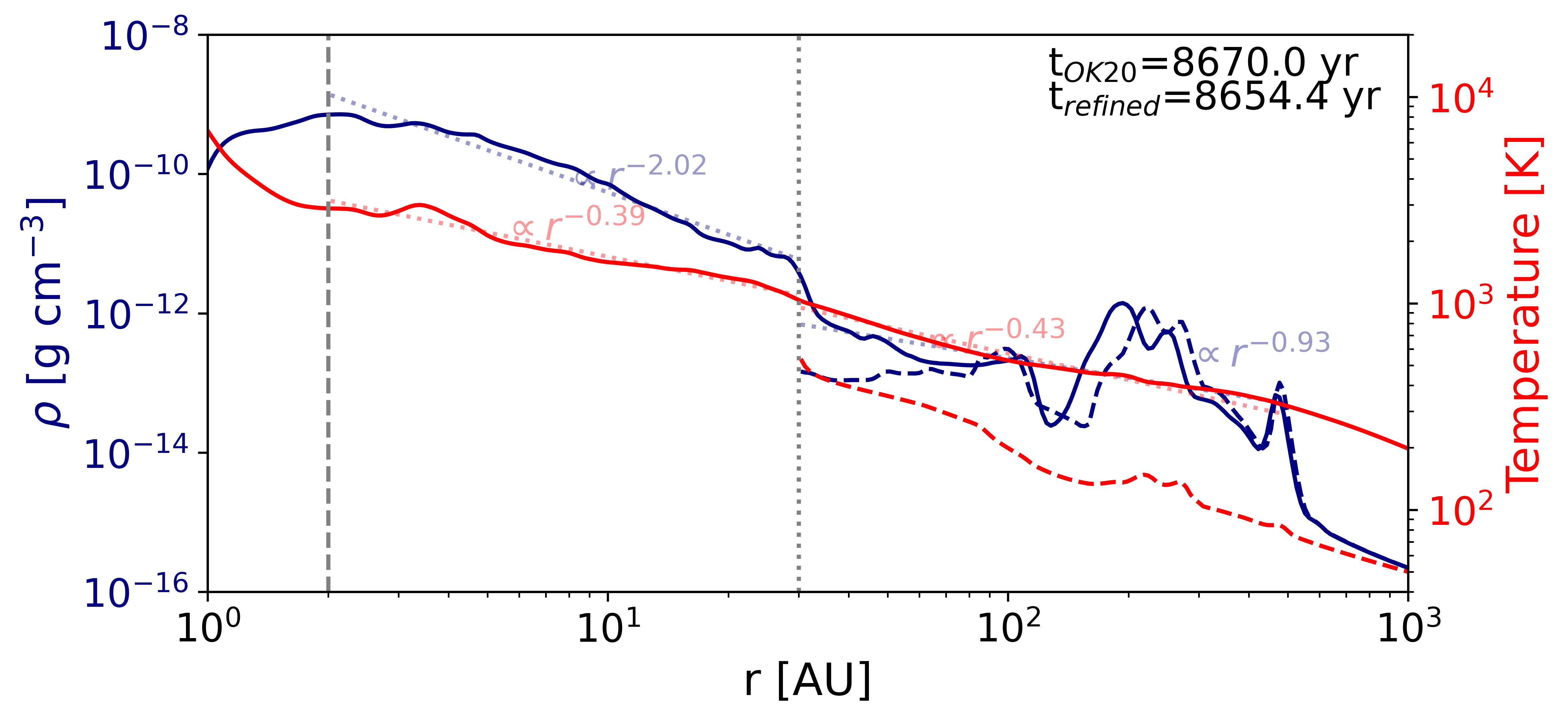}
    \caption{Azimuthally averaged midplane radial profiles of gas density (left y‑axis) and temperature (right y‑axis) for paired output times from the OK20 run (dashed) and the refined run (solid). Power‑law fits (dotted) are shown for two radial ranges: 2–30 AU and 30–500 AU, with fitted exponents annotated at representative radii. Vertical lines mark 2 AU and 30 AU, separating the inner fit region from the outer disk. The fitting procedure and fit parameters are reported in Appendix~\ref{sec:power_law}. Times corresponding to each run are indicated in the upper-right corner. The top two panels compare the two simulations at the same absolute times (starting from the restart at $t=8580$~yr), while the bottom two panels show profiles aligned to each model’s burst peak (i.e., each run’s peak $\dot{M}$ is set to $t=0$), isolating the post-burst thermodynamic response. }
    \label{fig:profiles}
\end{figure}

To understand the disk’s thermodynamic evolution, we examine the midplane structure of both models in detail. In Figure~\ref{fig:profiles}, we compare azimuthally averaged radial profiles of midplane gas density and temperature for the OK20 and refined runs in two complementary ways: first by inspecting both models at the same absolute times (starting from the restart at $t=8580$ yr), and second by comparing each model relative to its own accretion-rate peak (i.e. taking the moment of peak $\dot{M}$ in each model as $t=0$). Both approaches are informative but answer different questions. Direct comparison at the same absolute times shows how the two numerical setups evolve from an identical initial state and highlights differences in migration speed and timing; comparison relative to each model’s burst time removes the timing offset introduced by the different migration rates and therefore isolates the post-burst thermodynamic response. 

When we compare the two models at the same absolute times (starting at $t=8580$ yr), the dominant and most robust difference is in the temperature evolution at large radii (we use the midplane temperature at 100 AU as a convenient reference point). Both models begin at $T\approx270$ K at 100 AU at restart, but the refined model heats much faster: within the first 30 yr the midplane temperature at 100 AU grows from 270 K to values of order a few hundred Kelvin. Quantitatively, in successive 10-yr bins the refined run shows effective heating rates at 100 AU of order a few to a few tens of K~yr$^{-1}$ (e.g., $\sim4.6$ K~yr$^{-1}$ between 8580–8590 yr, $\sim13.4$ K~yr$^{-1}$ between 8590–8600 yr, and up to $\sim24$ K~yr$^{-1}$ between 8600–8610 yr). In the same intervals the OK20 run heats much more slowly (typically $\lesssim 1$ K~yr$^{-1}$ early on, rising to $\sim3-8$~K~yr$^{-1}$ closer to the burst). These numbers quantify the earlier statement that the refined model becomes hotter earlier.

The density profiles outside 30 AU remain broadly similar between the two runs: the global surface density and large-scale disk structure are not dramatically altered by the inner boundary treatment because most of the disk mass and gravitational torques lie at larger radii. The primary morphological difference is that the fragment (visible as a pronounced bump in the radial density profile) migrates inward faster in the refined model. That faster inward drift causes the density bump to sweep inward earlier, which in turn is the principal driver of the earlier temperature rise seen at 100 AU in the refined run.

The faster inward drift of the fragment in the refined run is primarily a dynamical consequence of resolving the inner few AU and the connecting spiral arm.  In the refined model the inner disk and spiral wakes are retained in the computational domain and can therefore exert stronger, coherently negative gravitational torques (Lindblad plus coorbital contributions) on the fragment and also remove angular momentum by advecting mass inward along the arm.  By contrast, the large 30~AU sink in OK20 removes this inner reservoir and damps the inner part of the wake, which reduces the net negative torque and effectively filtering migration %on a longer viscous timescale.  
into a more gradual, volume-averaged delivery. Secondary effects – slightly larger bound mass inside the Hill sphere in the refined run and reduced numerical damping at higher resolution – further accelerate migration, so that together these factors produce the systematically faster inward drift seen in the refined model.

Turning now to the behavior across the burst itself: when we compare each model relative to its own peak accretion moment, the density profiles at the time of peak are nearly identical, confirming that both runs dump comparable amounts of mass into the inner boundary. However, the temperature structures differ markedly: at their respective peaks (third panel in Fig.~\ref{fig:profiles}), the refined model shows substantially higher midplane temperatures at 100~AU ($\sim$893~K) than the OK20 model ($\sim$584~K). After the burst, both models cool, but with different time dependence: in the first 10 years after peak the OK20 disk cools rapidly (from $\sim$584 K to $\sim$373 K, an average $\sim -21$ K~yr$^{-1}$), while the refined model undergoes an initially still faster drop (from $\sim893$ K to $\sim539$ K, $\sim -35$ K~yr$^{-1}$). Between 10 and 30 years after the peak the OK20 disk continues to cool substantially (reaching $\sim198$ K at 100~AU by 30~yr after peak), whereas the refined model remains comparatively hot ( $\sim$528~K at 30~yr). Averaging over the entire 30-yr post-peak interval produces similar mean cooling rates in the two models ($\sim -12$ to $-13$ K~yr$^{-1}$), but the evolution is structured: OK20 cools more steadily and ends up much colder at late times, whereas the refined run experiences a large initial temperature drop and then stalls at an elevated temperature plateau.

The dichotomy is straightforwardly explained by the fate of the inner few AU: in the refined model a hot, dense inner disk forms and survives the peak, continuing to produce local dissipation (shocks and advective heating)
%viscous dissipation 
and reprocessed radiation that irradiates and thermally diffuses outward, so the outer disk (e.g., at 100 AU) remains hotter on multi-decade timescales. By contrast, the OK20 run does not resolve this inner reservoir — the mass and local dissipative heating that would otherwise be concentrated at a few AU are swallowed by the large 30~AU sink and are therefore not available to heat the resolved outer disk, which relaxes more rapidly toward the stellar-irradiation equilibrium and becomes much colder at late times. 
The persistently higher temperatures in the refined run increase the local sound speed and therefore raise the Toomre $Q$ parameter, making the outer disk less prone to gravitational instability and new fragment formation \citep{1964Toomre, 2001Gammie, 2016Kratter}.

\subsection{Inner-disk accretion energetics and observational implications}

The refined model explicitly resolves a compact, hot inner disk during the burst (temperatures exceed $10^4$ K locally), so it is important to quantify the radiative power that this region can in principle produce.  A useful order-of-magnitude estimate is the accretion luminosity, $L_{\rm acc}\simeq G M_* \dot M / R_*$ evaluated with $M_*\!\simeq\!4.8\,M_\odot$, a representative stellar/inner–disk radius $R_*\!\sim\!5\,R_\odot$, and the peak mass flux $\dot M\sim4\times10^{-2}\,M_\odot\,\mathrm{yr}^{-1}$.  This yields $L_{\rm acc}$ of order $10^{6}\,L_\odot$ — i.e. well above the Eddington luminosity for a $\sim4.8\,M_\odot$ star ($L_{\rm Edd}\sim1.6\times10^{5}\,L_\odot$).  Such a short, intense power release is physically plausible for a TDE (tidal disruption event)-like event produced by fragment disruption, but the gross number alone does not tell us what an observer would actually see.

Two important caveats follow directly from the super-Eddington estimate: much of the locally liberated accretion energy can be trapped by the large optical depth \citep[e.g.,][]{2013KuiperYorke_opacity}, advected into the star, or channelled into mechanical outflows \citep[e.g.,][]{2015Kuiper, 2016Kuiper} rather than escaping as isotropic radiation, and radiation pressure (including line driving) at these luminosities will likely launch strong winds that alter the inner–disk structure \citep[e.g.,][]{2019KeeKuiper} and its thermal coupling to the outer disk. Both effects substantially reduce the fraction of $L_{\rm acc}$ that is available to heat the resolved outer disk and to produce observable bolometric flux, so simple energetics alone cannot predict the observational appearance. From an observational perspective, this is exactly where the refined and OK20 models should diverge most: the refined run contains the forming inner disk – luminous source that can directly irradiate and reprocess energy in the surrounding disk, whereas the OK20 setup effectively removes that source by absorbing inner dissipation into a large sink and therefore underpredicts any direct or reprocessed emission from the inner few AU. To quantify these differences and predict observables we perform detailed radiative post-processing using RADMC-3D \citep{2012Dullemond}. The results of those calculations are presented in the next section.

%%%%%%%%%%%%%%%%%%%%%%%%%%%%%%%%%%%%%%%%%%%%%%%%%%%%%%%%%%%%%%%%%%%%%%%%%%
%%%%%%%%%%%%%%%%%%%%%%%%%%%%%%%%%%%%%%%%%%%%%%%%%%%%%%%%%%%%%%%%%%%%%%%%%%

\section{Observational Consequences of Fragment Disruption and Accretion Bursts}\label{sec:radmc}

\begin{figure}
    \centering
    \includegraphics[width=1\linewidth]{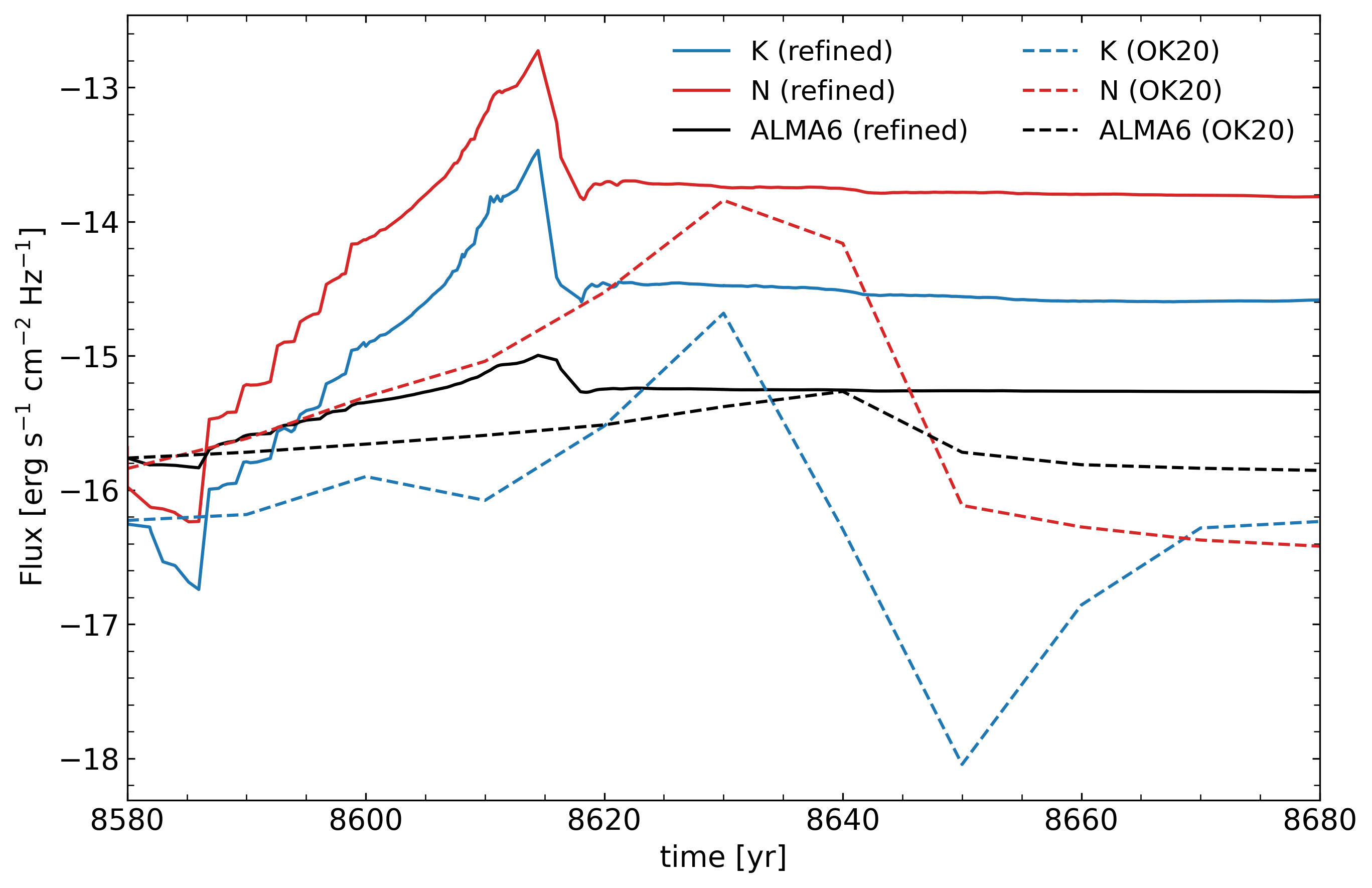}
    \caption{Time evolution of band flux densities from RADMC-3D post-processing for two models during an outburst. Shown are K (blue), N (red), and ALMA Band 6 (black) flux densities versus time. Solid lines denote the refined model and dashed lines the OK20 model. 
    }
    \label{fig:lightcurves}
\end{figure}

\begin{figure}
    \centering
    \includegraphics[width=1\linewidth]{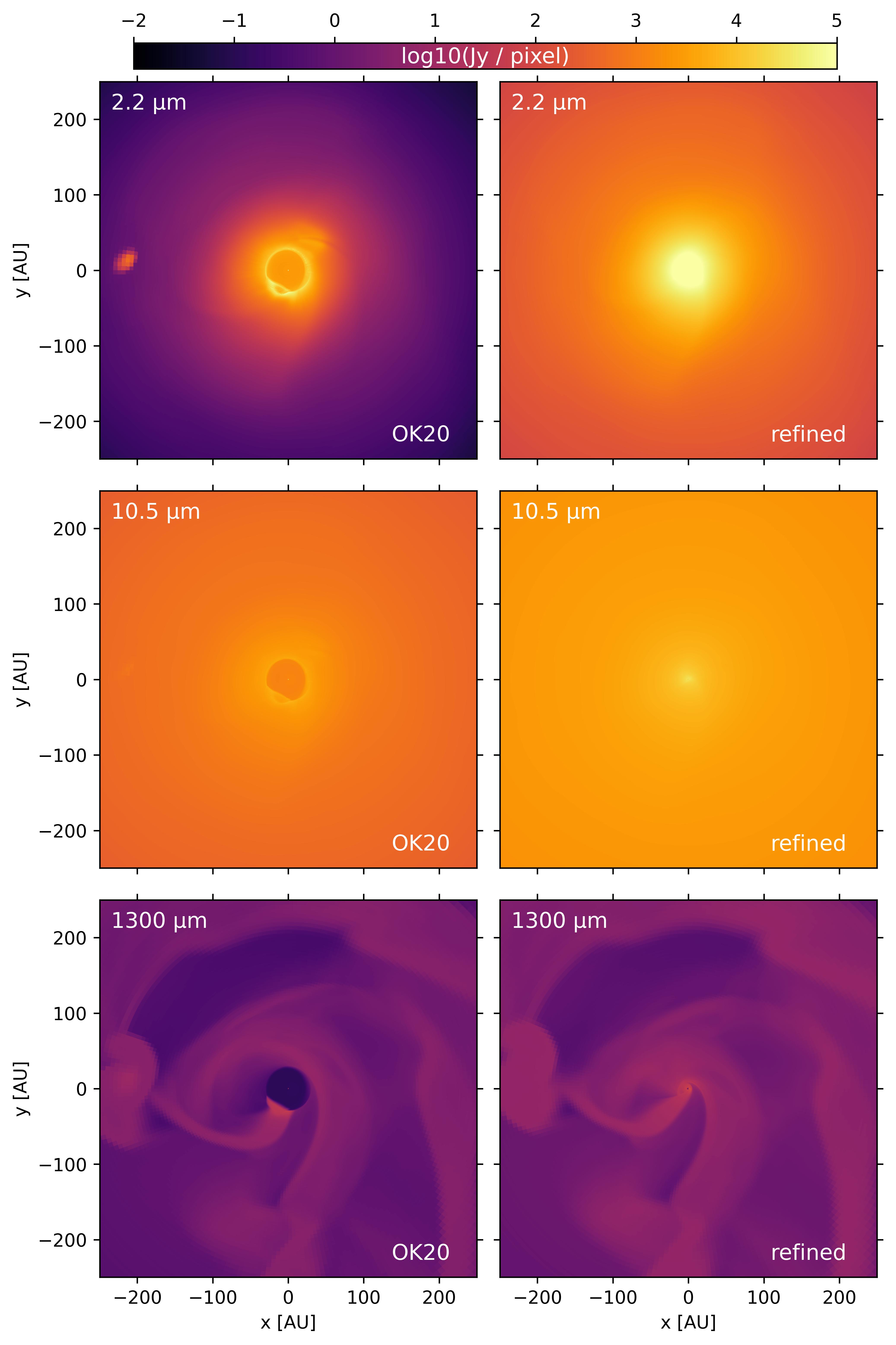}
    \caption{Images from RADMC-3D for the two models (OK20 on left, refined on right). From top to bottom: K, N, and ALMA Band 6. $t_{\rm OK20}=8630$~yr, $t_{\rm refined}=8614.4$~yr (peak of the burst in both models)}
    \label{fig:bands_images}
\end{figure}

To translate the differences between the OK20 and refined runs into observables, we post-processed matched hydrodynamic snapshots with RADMC-3D and extracted time series fluxes in three representative bands at an assumed source distance of 1~pc: near-IR (K), mid-IR (N) and millimeter (ALMA Band 6). Figure~\ref{fig:lightcurves} shows the temporal evolution of the band fluxes. 

Both models share similar pre-burst baselines around the restart time ($t=8580$~yr). However, the refined model produces an outburst that is an order of magnitude more luminous and morphologically distinct from OK20. In the refined model, the K and N fluxes peak synchronously at $t\approx8614.45$~yr (reaching dynamic ranges of $\sim$600 and $\sim$1800, respectively), followed by a prolonged IR afterglow. In contrast, OK20 produces a much weaker IR peak (dynamic ranges $\sim$35 and $\sim$100 at $t\approx8630$~yr) that decays rapidly. The millimeter response in OK20 is also noticeably delayed ($\sim$10~yr lag) compared to the IR, consistent with heating propagating from the sink boundary to the cooler extended dust, while ALMA6 brightens more synchronously in the refined run.

Examining band ratios at the IR maxima clarifies these spectral differences: the refined model concentrates a much larger fraction of the emergent power into warm/hot dust ($N/\mathrm{ALMA6} \sim$186 vs $\sim$34 for OK20). This arises directly from the underlying hydrodynamics (Sect.~\ref{sec:therma_structure}): the refined run forms a compact, hot inner disk that irradiates efficiently, whereas OK20 absorbs inner-disk dissipation into the unmodeled sink region.

To make the physical contrasts immediately apparent, Figure~\ref{fig:bands_images} presents matched RADMC-3D images at the IR maxima. At millimetre wavelengths (bottom), the models appear broadly similar because ALMA6 traces cool, extended disk substructure (giving comparable peak fluxes). However, at short wavelengths, the difference is stark. In all OK20 panels, a pronounced low-flux central cavity is evident. By contrast, the refined images reveal a compact, luminous inner source that completely dominates the short-wavelength emission.

The K-band comparison yields a particularly instructive observational consequence. In OK20 a secondary fragment at $\sim$200~AU appears as a distinct bright spot because the central region is faint in the resolved image. In the refined model that same fragment is effectively invisible at K because the intense inner disk emission overwhelms and masks it. Thus, two simulations that inject comparable total mass into the inner region can produce qualitatively different observability: the refined configuration predicts bursts dominated by a compact IR core that can hide more distant compact sources, while OK20-like setups, with a large unresolved sink, permit outer fragments to remain visible in the near-IR.

These contrasts have clear observational consequences. In nature, the inner disk is always present. However, different burst mechanisms imprint their energy on different spatial scales. Bursts associated with compact inner-disk heating are expected to produce strong, rapidly rising NIR–MIR emission from a small region close to the star and a prolonged IR high state, whereas bursts fed by more extended, diffuse inflow should yield weaker short-wavelength brightening, a stronger response at longer wavelengths, and a delayed mm excess tracing larger disk radii. Observations then modulate these intrinsic signatures through instrumental resolution and sensitivity: facilities such as JWST and ground-based MIR instruments preferentially probe the compact inner regions, while mm interferometers trace the cooler, more extended disk\footnote{In practice, ALMA on typical baselines may not distinguish the two scenarios.}

These radiative diagnostics therefore reflect physical differences in mass deposition and dissipation, rather than modeling choices such as sink prescriptions. Nevertheless, whether and how these differences can be identified observationally depends on the spatial scales accessible to a given instrument. Identical mass-delivery events can thus appear qualitatively different in the data, not because the inner disk is absent, but because different observations emphasize different parts of the disk where the burst energy is released.

%%%%%%%%%%%%%%%%%%%%%%%%%%%%%%%%%%%%%%%%%%%%%%%%%%%%%%%%%%%%%%%%%%%%%%%%%%%%%%%%%%%%

%%%%%%%%%%%%%%%%%%%%%%%%%%%%%%%%%%%%%%%%%%%%%%%%%%%%%%%%%%%%%%%%%%%%%%%%%%
%%%%%%%%%%%%%%%%%%%%%%%%%%%%%%%%%%%%%%%%%%%%%%%%%%%%%%%%%%%%%%%%%%%%%%%%%%

\section{Formation, migration, and survival of second Larson core} \label{sec:second_core}

The formation of stellar companions via disk fragmentation critically depends on the thermodynamic transition from a collapsing gas fragment’s first hydrostatic core to its second Larson core (SC). Early 1D radiation‐hydrodynamic studies demonstrated that molecular hydrogen dissociation at $\sim$2000~K triggers rapid core formation and sets its initial mass and radius \citep{2000MasunagaInutsuka}. Subsequent multidimensional simulations have explored the interplay of rotation, magnetic fields, and radiative feedback in shaping second‐core properties \citep{2015Tsukamoto, 2017Vaytet, 2020Bhandare}. In massive protostellar disks, these nested cores may survive long enough to become bound companions, but their fate depends on inward migration, tidal disruption, and contraction. In this section, we first compute the orbital trajectory of a newly formed second core around a 4.8$\msun$ star (mass of the star in our models), then apply Roche‐limit and Kelvin–Helmholtz analyses to quantify its survivability against tidal disruption as it spirals inward.

The formation of stellar companions around massive stars involves complex dynamical processes that depend critically on the thermal evolution of collapsing gas fragments. When the central temperature of a gravitationally bound fragment exceeds the dissociation threshold of molecular hydrogen ($\sim$2000 K), the H$_2$ molecules begin to dissociate, triggering the formation of a SC \citep{1969Larson, 2000MasunagaInutsuka}. This process marks a fundamental transition from the optically thin first collapse phase to the formation of an optically thick hydrostatic protostellar core, which will eventually become a stellar companion.

\subsection{Second–core trajectory}
To explore whether a compact SC inferred from our snapshot could survive and migrate inward under the influence of the central star, we performed illustrative point-mass integrations starting from the momentum-weighted center-of-mass position and velocity of the candidate core. The integrations include the evolving stellar mass (from the OK20 accretion history) and span 10~kyr with 0.01-yr timesteps, but they neglect disk self-gravity, hydrodynamic drag, multi-body encounters and gas torques; therefore the results are intended only as qualitative indicators rather than quantitatively robust predictions (full method, assumptions and limitations are given in Appendix~\ref{sec:trajectory}).

Under these assumptions, a compact SC with $M_{\rm sc}\simeq3.6\times10^{-2}\ \msun$ remains bound and follows a highly eccentric, inward-spiraling orbit. The computed periastron in the integration is $a_{\rm peri}\approx17$~AU, the late-time eccentricity approaches $e\approx0.55$, and the periastron decline rate is $\dot a_{\rm peri}\simeq-4.94\times10^{-3}$~AU\,yr$^{-1}$. Near the end of the 10 kyr integration the instantaneous orbital period is $\sim50$~yr, and a simple extrapolation of the point-mass trajectory gives a nominal time-to-impact of order $1.35\times10^{4}$~yr.

Periodic or quasi-periodic variability on comparable (year–to–decade) timescales has been reported in several massive and intermediate-mass young stellar objects in long-term infrared and maser monitoring campaigns. In some cases, bursts separated by many years may observationally appear as distinct events. A bound secondary on a decadal orbit could naturally produce such signatures by periodically modulating the inner-disk accretion flow or irradiation geometry. A brief overview of relevant observational examples is provided in Appendix~\ref{sec:trajectory}.

\subsection{Tidal survivability and contraction of the second core} \label{sec:sc_contraction}

To determine whether the newly formed SC can survive its inward migration under the tidal field of the central star, we evaluate two complementary criteria: the Roche‐limit criterion and the Kelvin–Helmholtz contraction timescale.

Based on the 1D collapse models of \citet{2020Bhandare}, interpolating between their 0.5$\msun$ and 1.0$\msun$ cases for our 0.61$\msun$ first core yields $M_{\rm sc}\simeq3.6\times10^{-2}\msun,\;
R_{\rm sc} \simeq 0.214\;\mathrm{AU},\; \dot M_{\rm sc}\simeq4\times10^{-5}\msunyr,$ with a central temperature $T_{\rm sc}\simeq2\times10^3$~K .

The Roche radius for a companion of mass $M_{\rm sc}$ orbiting a star of mass $M_*$ at separation $a$ is given by
\begin{equation} 
    R_{\rm Roche}(a)\approx a \left( \frac{M_{\rm sc}}{3\,M_*} \right) ^{1/3}.
\end{equation}
For $M_*=4.8\msun$ and the periastron after 10 kyr, $a_{\rm peri}=17$~AU, we have $ R_{\rm Roche}(a_{\rm peri})\approx2$~AU, which greatly exceeds $R_{\rm sc}$.  Tidal disruption first occurs when $R_{\rm Roche}\approx R_{\rm sc}$; solving for the critical separation
\begin{equation}
    a_{\rm disrupt}\simeq R_{\rm sc}\, \left( \frac{3\,M_*}{M_{\rm sc}} \right) ^{1/3}
    \approx1.7\;\mathrm{AU}.   
\end{equation}
Using the measured periastron‐decline rate $\dot a_{\rm peri}\approx-4.94\times10^{-3}$~AU~yr$^{-1}$, the time to reach $a_{\rm disrupt}$ is $\sim3.1\times10^3$~yr.

While migrating inward, the SC continues to accrete and radiate away gravitational energy, contracting on its Kelvin-Helmholtz timescale
\begin{equation} 
    t_{\rm KH}\simeq\frac{G\,M_{\rm sc}^2}{R_{\rm sc}\,L_{\rm sc}},
\end{equation}
where we approximate the core’s luminosity by its accretion luminosity
\begin{equation}
    L_{\rm acc}\simeq\frac{G\,M_{\rm sc}\,\dot M_{\rm sc}}{R_{\rm sc}}\sim1\,\lsun.
\end{equation}
Substituting numbers gives $t_{\rm KH}\sim9\times10^2$~yr, a few times shorter than the time to reach $a_{\rm disrupt}$. The SC therefore contracts by factors of a few well before the nominal disruption radius is reached, so its radius shrinks faster than the orbit decays and tidal disruption is postponed—or avoided—altogether. As a result, second cores formed by disk fragmentation can survive to much smaller radii (a few AU) before being violently stripped; when disruption does occur it deposits material into the inner disk on a dynamical timescale and can trigger a luminous accretion outburst like those observed in high-mass young stellar objects (HMYSOs) \citep[e.g.,][]{2017Caratti}.

These results directly connect to the scenario explored in our recent works on episodic accretion in HMYSOs \citep{2021ElbakyanNayakshin, 2023ElbakyanNayakshin}, where tidal disruption of post‐collapse fragments or massive gaseous protoplanets was proposed as a viable mechanism for generating bursts. While 1D disk models capture the inner‐disk thermal instabilities and magnetorotational effects, they are inherently limited in describing non‐axisymmetric fragment migration and disruption. The trajectory analysis presented here bridges this gap, demonstrating that second cores formed via disk fragmentation are dynamically robust enough to migrate deep into the star’s tidal sphere, where their eventual disruption can provide the mass deposition needed to produce FU Ori–like outbursts in HMYSOs.

%%%%%%%%%%%%%%%%%%%%%%%%%%%%%%%%%%%%%%%%%%%%%%%%%%%%%%%%%%%%%%%%%%%%%%%%%%%%%%
%%%%%%%%%%%%%%%%%%%%%%%%%%%%%%%%%%%%%%%%%%%%%%%%%%%%%%%%%%%%%%%%%%%%%%%%%%%%%%
%%%%%%%%%%%%%%%%%%%%%%%%%%%%%%%%%%%%%%%%%%%%%%%%%%%%%%%%%%%%%%%%%%%%%%%%%%%%%%
\section{Comparison with observed bursts on intermediate / massive protostars}

Accretion bursts have been observed in several high- and intermediate-mass young stellar objects, with a range of amplitudes and timescales. A particularly informative case is M17 MIR, an intermediate-mass protostar $M\simeq5.4\msun$ (stellar mass is $4.8\msun$ in our models) that shows repeated mid-IR brightening episodes of order $\sim$2 mag and durations of order 9–20 yr \citep[two bursts separated by a few years of quiescence;][]{2021Chen}.

Our radiative-transfer light curves show that a diffuse fragment being sheared and accreted through a resolved inner spiral can produce long-duration increases in IR flux that — if one defines the burst only above some detection threshold — can have apparent durations comparable to the decade-long events reported for M17 MIR. In this sense the total duration and integrated energy of a fragment-driven episode can be similar to observed decade-scale bursts. However, an important mismatch remains in the rise time: the best-observed massive YSO bursts (and many well-studied cases such as S255IR-NIRS3 and others) show very rapid rises, often $\ll$1~yr, whereas our fragment–accretion light curves rise much more gradually (tens of years to reach peak). 
This slow ramp is an inherent dynamical outcome of a diffuse clump feeding the inner disk through an extended spiral stream: mass is delivered progressively rather than impulsively, and the buildup of material in the inner few AU proceeds on orbital/advective timescales set by this extended geometry.

These facts lead to a simple physical conclusion: while the accretion of diffuse fragment material can plausibly explain long-duration outbursts, it is unlikely to produce the very short rise times observed in many HMYSO bursts. By contrast, the compact second Larson cores that form at the centres of collapsing fragments are physically much smaller and denser. These compact objects can survive tidal torques while migrating much closer to the star before being disrupted; a disruption at small radii deposits mass and energy on short dynamical timescales, naturally producing a fast rise and a high-amplitude, short-timescale flare. Our interpolation of 1D collapse models and the estimated Kelvin–Helmholtz contraction times (see Sect.~\ref{sec:sc_contraction}) indicate that second cores will contract rapidly and become more resistant to early disruption, making close-in disruption (and hence rapid bursts) plausibly common for those objects.

Observationally, this hypothesis suggests two testable diagnostics. First, bursts produced by compact-object disruption should be proportionally stronger in the near/mid-IR and show very rapid rises, with compact, high-brightness inner emission in high-resolution IR imaging (and possibly prompt high-velocity ejection signatures), whereas diffuse-fragment-driven bursts should show more gradual rises, stronger long-wavelength (mm) response and clearer extended spiral/stream morphology. Second, time-domain spectroscopy and maser kinematics (e.g., water masers) during bursts can constrain where the mass was deposited: close-in impulsive disruptions should produce signatures of rapid inner-disk heating or inner-envelope ejection, while diffuse feeding will give slower, spatially extended responses. Observational strategies that combine high-cadence IR monitoring, high-resolution imaging (JWST and ALMA), and maser/line kinematics will therefore be decisive in distinguishing between these scenarios.

%%%%%%%%%%%%%%%%%%%%%%%%%%%%%%%%%%%%%%%%%%%%%%%%%%%%%%%%%%%%%%%%%%%%%%%%%%%%%%
%%%%%%%%%%%%%%%%%%%%%%%%%%%%%%%%%%%%%%%%%%%%%%%%%%%%%%%%%%%%%%%%%%%%%%%%%%%%%%
%%%%%%%%%%%%%%%%%%%%%%%%%%%%%%%%%%%%%%%%%%%%%%%%%%%%%%%%%%%%%%%%%%%%%%%%%%%%%%
\section{Conclusions}

We have followed the evolution of a gravitationally bound fragment in a massive protostellar disk from pre-collapse through inward migration, tidal disruption and the resulting accretion burst, comparing an original large-sink simulation from \citep[][OK20 model]{2020OlivaKuiper} with a new high-resolution run that resolves the inner few AU (the refined model). The main conclusions are the following.

-- Despite nearly identical peak accretion rates, the refined model produces a somewhat shorter burst (FWHM 4.7~yr vs 6.7~yr) with a faster rise and a noticeably sharper immediate post-peak decline. This reflects that, in the refined run, mass is deposited into a compact inner disk and drained on shorter, more variable timescales.
By contrast, the OK20 run — with its larger (30~AU) sink — spreads the accretion into a longer, smoother episode. This behavior reflects where the mass is lost: in OK20 the fragment passes through the large sink (losing only part of its mass) and is removed from the resolved domain, so the recorded accretion is effectively averaged over a large, unresolved volume and appears smoother and more extended. In the refined run the fragment is tidally disrupted inside the resolved inner few AU and its material collects in a compact inner disk; that locally concentrated reservoir drains and falls back on shorter, more variable timescales, producing the briefer event with a steeper initial decline.

-- The fragment’s center crosses the H$_2$ dissociation threshold $\sim2000$~K, consistent with imminent second-Larson-core formation in collapse models. Interpolating 1D collapse results yields a plausible second-core mass of order $M_{\rm sc}\simeq3.6\times10^{-2}\msun$ and a compact radius $\sim0.2$~AU. Its Kelvin–Helmholtz contraction time $\sim10^3$~yr is shorter than the nominal time to Roche disruption in our simple estimate, implying that a contracted core would be more resistant to tidal disruption if it formed.

-- RADMC-3D post-processing reveals strong, observationally relevant contrasts. The refined model produces far larger near– and mid-IR peaks and a long, elevated IR tail; ALMA Band-6 brightening is more modest and similar in both runs because mm emission traces cooler, extended dust. Crucially, a luminous compact inner disk in the refined case can outshine and mask more distant fragments at short wavelengths, so the inner disk’s configuration and emission materially affects which structures are observable.

-- Accretion of diffuse fragments can deliver sufficient mass and total radiated energy to produce outbursts that last decades, comparable in integrated strength to some HMYSO events such as M17 MIR. However, the slow rise time we obtain contrasts with the very short $\ll1$~yr rises seen in some HMYSO bursts; additionally, a number of reported HMYSO events have overall durations of order a few years or less, which are difficult to reconcile with diffuse-fragment infall. This points again toward compact second-core disruptions — which can survive to much smaller radii and deposit mass on short dynamical timescales — as the more plausible origin for the fastest, highest-amplitude flares.

-- This persistent inner heating raises the Toomre $Q$ at tens–hundreds of AU and can suppress further fragmentation, reducing the likelihood of new companion formation.

Numerically resolving the inner few AU in fragmenting massive disks changes where and how fragment mass is deposited, alters burst morphology, and substantially modifies observable photometric and morphological signatures. Predicting realistic episodic accretion outcomes for massive protostars therefore requires inner-disk resolution combined with radiative, self-gravitating, multi-physics calculations.

\begin{acknowledgements}
We thank the anonymous referee for their constructive review, which significantly improved the quality and clarity of this paper. RK acknowledges financial support via the Heisenberg Research Grant funded by the Deutsche Forschungsgemeinschaft (DFG, German Research Foundation) under grant no.~KU 2849/9, project no.~445783058. The authors gratefully acknowledge the computing time granted by the Center for Computational Sciences
and Simulation (CCSS) of the University of Duisburg-Essen and provided on the cluster amplitUDE (DFG Projects 459398823 / grants INST 20876/423-1 FUGG) at the Zentrum für Informations- und Mediendienste (ZIM).
\end{acknowledgements}


\bibliographystyle{aa}
\bibliography{ref_base2}
% \bibliography{references}

\begin{appendix}

\section{Physics of the numerical model} \label{sec:model}

Here, we briefly describe the main properties of our model, while more details can be found in \citetalias{2020OlivaKuiper}. Our simulations model the three-dimensional evolution of an astrophysical system, incorporating self-gravity, hydrodynamics, and radiation transport. The system is treated as an ideal gas, with its dynamics governed by the standard hydrodynamics equations:  
\begin{equation}
    \frac{\partial \rho}{\partial t} + \nabla \cdot (\rho \mathbf{u}) = 0
\end{equation}
\begin{equation}
    \frac{\partial}{\partial t} (\rho \mathbf{u}) + \nabla \cdot (\rho \mathbf{u} \otimes \mathbf{u} + P \mathbf{I}) = \rho \mathbf{a}_{\text{ext}}
\end{equation}
\begin{equation}
    \frac{\partial E}{\partial t} + \nabla \cdot ((E + P) \mathbf{u}) = \rho \mathbf{u} \cdot \mathbf{a}_{\text{ext}}
\end{equation}
where $\rho$ is the density, $\mathbf{u}$ the velocity field, $P$ the pressure, and $E$ the total energy density. The external acceleration term, $\mathbf{a}_{\text{ext}}$, accounts for gravitational forces from the central object ($\mathbf{a}_*$), self-gravity ($\mathbf{a}_{\text{sg}}$), and radiation forces ($\mathbf{a}_{\text{rad}}$). These equations are solved using the hydrodynamics module of the numerical grid code Pluto \citep{2007Mignone}, with additional modules for self-gravity \citep{2010Kuiper_a} and radiation transport \citep{2010Kuiper_b, 2020Kuiper}.

Self-gravity is solved using the Poisson equation:  
\begin{equation}
    \nabla^2 \Phi_{\text{sg}} = 4\pi G \rho
\end{equation}
where $\Phi_{\text{sg}}$ is the gravitational potential. The gas follows a calorically perfect equation of state,  
\begin{equation}
    P = (\gamma - 1) E_{\text{int}}
\end{equation}
where $E_{\rm int}$ is the internal energy density and $\gamma$ the adiabatic index. Radiation transport is handled using a two-temperature flux-limited diffusion (FLD) approach  \citep{2020Kuiper}, 
complemented by a frequency-dependent stellar irradiation module identical to that used in \citetalias{2020OlivaKuiper}. The irradiation field is computed using \citet{1993Laor} dust opacities together with a constant gas opacity (0.01~cm$^2$~g$^{-1}$), and dust sublimation is treated following \citet{2005IsellaNatta}. The stellar luminosity and radius evolve according to accreting protostar tracks from \citet{2009HosokawaOmukai}, providing a realistic irradiation source term.

\section{Inner-grid extrapolation} \label{sec:extrapol_appendix}

\begin{figure}
    \centering
    \includegraphics[width=1\linewidth]{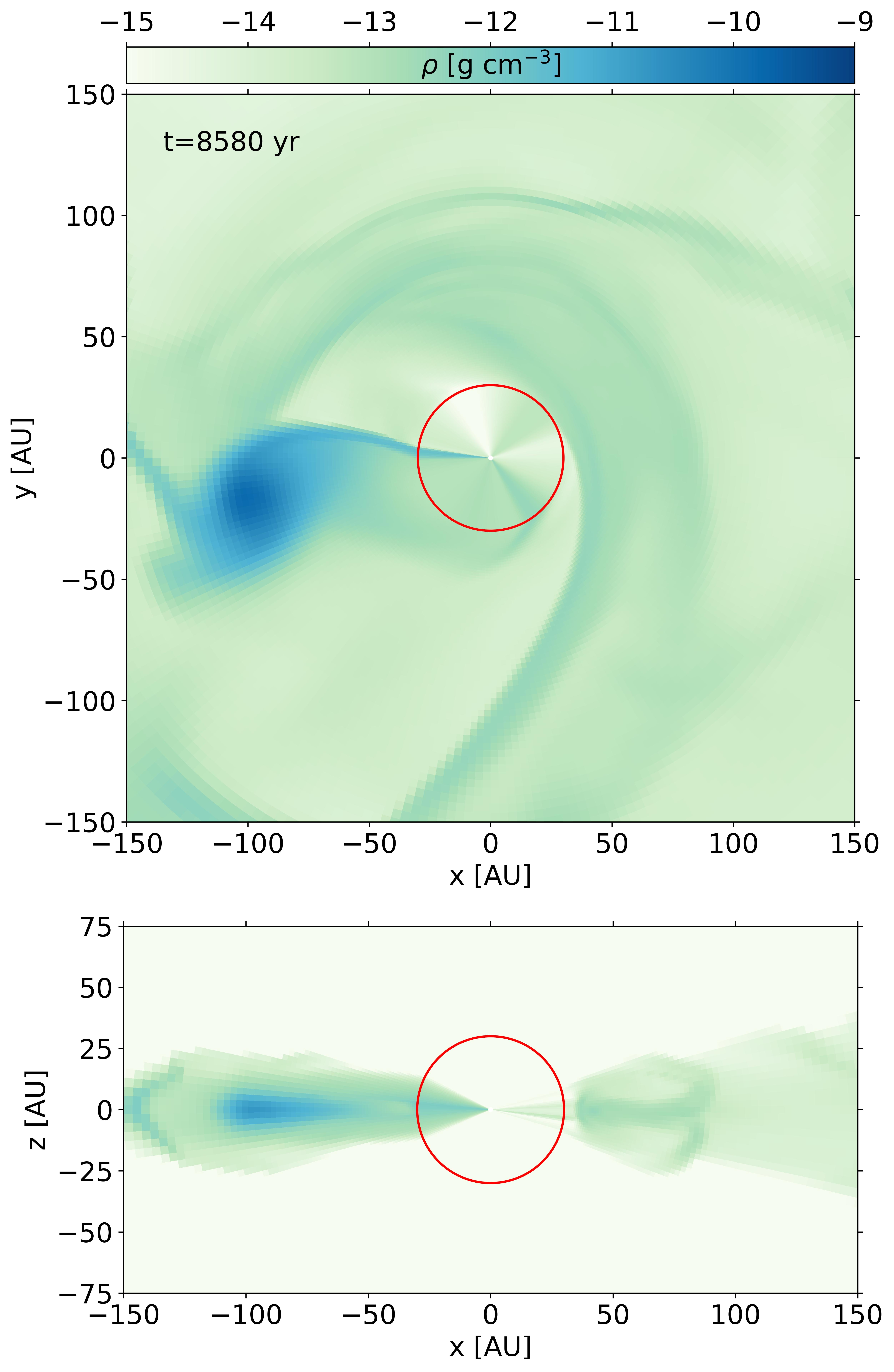}
    \caption{Initial gas mass density distribution in x-y (top) and x-z (bottom) planes. The red circle shows the size of the extrapolated region, which was initially considered as a sink cell in \citetalias{2020OlivaKuiper}.}
    \label{fig:rho_init}
\end{figure}

To initialize the refined radial domain (inner boundary moved from 30~AU to 1~AU) we populated the added grid cells by extrapolating the hydrodynamic variables (e.g., density, temperature, velocity) from the OK20 snapshot. Extrapolation was performed using the SciPy Interpolation sub-package \citep{2020SciPy-NMeth}. Specifically, for each variable we first extended the radial grid inward by adding $N_\mathrm{new}=140$ radial cells, constructed with the same constant geometric spacing factor as in the original grid (i.e., using the ratio $b=r_1/r_0$ so that the new radii follow $r_i=a\,b^i$ and smoothly continue the original logarithmic-like spacing), which places the new inner boundary at $\simeq 1$~AU. We then initialized the added cells by nearest-neighbor assignment using \texttt{scipy.interpolate.griddata} with \texttt{method='nearest'}: the new radial cells were marked as undefined and subsequently filled by copying the value from the closest existing cell of the original snapshot (in practice, this means the added inner region inherits the angular structure of the innermost resolved OK20 cells without imposing any functional form such as a linear or power-law extrapolation).
This method assigns values to the new grid points based on the nearest available data from the original simulation, ensuring a smooth transition between the previously resolved and newly introduced regions. The resulting initial density distribution is shown in Figure~\ref{fig:rho_init}, where the top and bottom panels display the x–y and x–z planes, respectively, in the inner part of the grid. The red circle marks the extent of the extrapolated region, which corresponded to the sink cell in \citetalias{2020OlivaKuiper}. 
We emphasize that this procedure is not intended to enforce an exact hydrostatic equilibrium inside 30~AU; rather, it provides a well-defined, reproducible initial guess that avoids undefined states in the newly introduced cells.
During the first few output time snapshots after restart, the extrapolated inner region naturally relaxes and adjusts to the surrounding disk, quickly settling into a state consistent with the original structure beyond 30 AU. We note that each output interval corresponds to many (typically hundreds to $\sim$thousand) hydrodynamical timesteps, so the relaxation occurs over a physically meaningful evolution time rather than over only a handful of numerical steps.

\section{Fragment identification and characterization methodology} \label{sec:fragment_contour}

\begin{figure}
    \centering
    \includegraphics[width=1\linewidth]{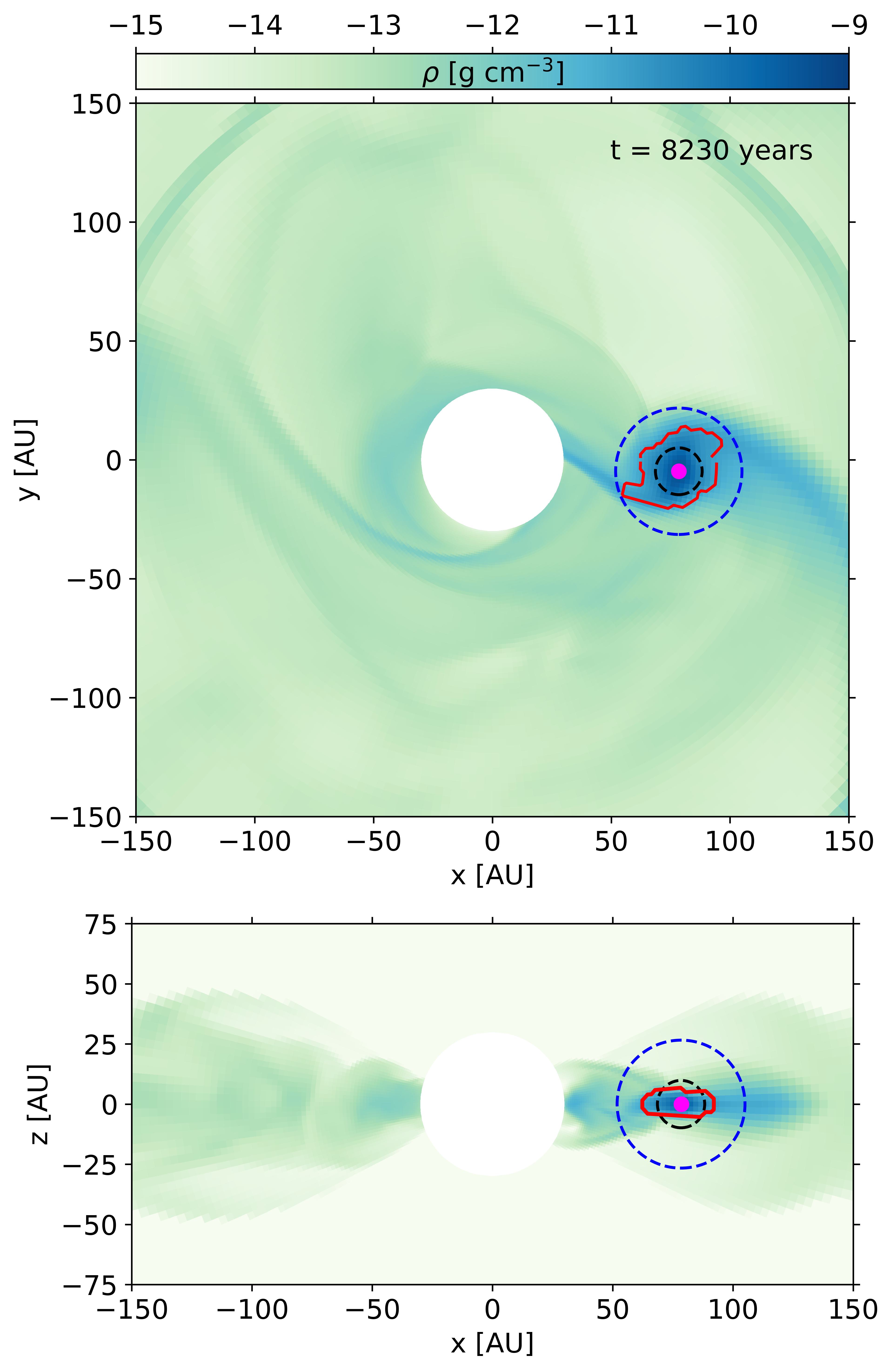}
    \caption{Figure showing the fragment properties in the disk. The colour map displays the gas surface density, the magenta dot marks the density peak, and the red contour outlines the 3D fragment surface defined by our temperature–density criterion. The black dashed circle indicates the fragment’s characteristic radius. The blue dashed circle shows the Hill radius.}
    \label{fig:fragment_contour}
\end{figure}

Figure~\ref{fig:fragment_contour} illustrates our fragment identification methodology through a representative example, displaying the gas mass density distribution in both the disk midplane (upper panel) and a vertical cross-section at $y=0$ (lower panel), which together reveal the three-dimensional morphology and structure of the identified fragments. The identification and characterization of gravitationally bound fragments in our disk fragmentation simulations ensures that identified objects represent physically meaningful condensations rather than transient density fluctuations.

Our methodology begins by identifying regions of elevated gas temperature as potential sites of gravitational collapse. We apply an initial temperature threshold of 700~K to the gas temperature field in the disk midplane, which effectively captures regions where compressional heating due to gravitational collapse becomes significant, distinguishing them from the cooler background disk material. This temperature criterion provides the initial boundaries for potential fragment regions and serves as a first-order filter for identifying thermally active sites.

Within each temperature-defined region, we locate the position of maximum density to determine the gravitational center of each potential fragment. This search is constrained to a specific radial range in the disk to focus on the region where fragmentation is expected while excluding both the central stellar environment and the outer disk regions where densities are insufficient for fragment formation. The maximum density location serves as the reference point for all subsequent analyses and is marked with a magenta point in both panels of Figure~\ref{fig:fragment_contour}.

To account for the local thermal environment and avoid artificially truncating fragment boundaries, we calculate an azimuthally averaged temperature at the radial distance corresponding to the density peak. This averaging is performed both in the full three-dimensional domain and specifically within the disk midplane to capture any vertical temperature gradients that might influence the fragment structure. We then define an adaptive temperature threshold by multiplying the maximum of these two averaged values by a factor of 1.3, ensuring that the fragment boundary encompasses material that is thermally coupled to the central condensation while excluding cooler, dynamically unrelated material.

Using this adaptive temperature threshold, we define the complete three-dimensional boundary of each fragment by identifying all regions where the gas temperature exceeds the calculated threshold within the specified radial constraints. The resulting fragment boundaries are visualized as red contours in Figure~\ref{fig:fragment_contour}.

Within the defined fragment volume, we calculate key physical properties including the total mass, characteristic radius and Hill radius. We calculate the fragment’s characteristic radius by measuring the distance from its density peak to every position on the defined surface and then taking the average of those distances. The Hill radius calculation incorporates the orbital radius of the density peak, the fragment mass, and the instantaneous stellar mass to assess the gravitational influence of each fragment relative to the central star. The Hill radius is represented as the blue dashed circle in Figure~\ref{fig:fragment_contour}, while the black dashed circle indicates the characteristic radius of the fragment. We note that the characteristic radius is noticeably smaller than the projected clump contour in the x-y plane because the fragment is vertically flattened, as seen in the x-z plane.

This methodology ensures that identified fragments represent physically meaningful, thermally distinct structures while adapting to the local thermal environment of each potential fragmentation site. The combination of temperature and density criteria, along with the adaptive threshold approach, helps distinguish genuine gravitational condensations from transient thermal fluctuations in the evolving disk.

\section{Second core trajectory analysis} \label{sec:trajectory}

\begin{figure}
    \centering
    \includegraphics[width=1\linewidth]{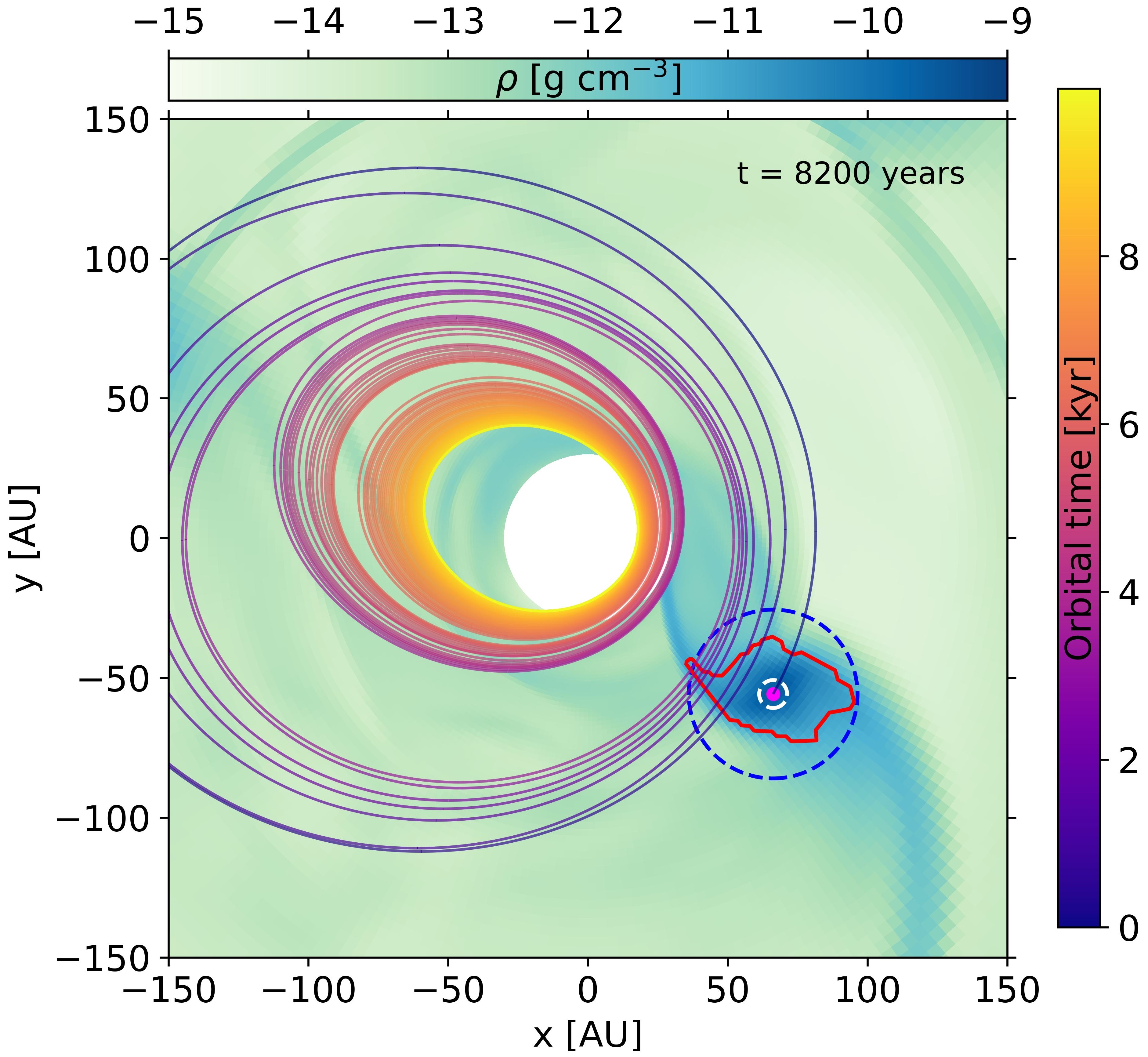}
    \caption{Trajectory of a gravitationally bound second Larson core (SC). The color-coded curve traces the computed orbital path over a 10-kyr integration, with color progression indicating temporal evolution. The initial velocity is determined via momentum-weighted averaging within a 5~AU sphere (white dashed line) centered on the fragment’s density maximum (magenta point). The red contour delineates the fragment boundary based on temperature-density criteria and the blue dashed circle marks the Hill radius. The background displays the midplane gas density. The highly eccentric orbit ($e\approx0.55$) and declining periastron illustrate the rapid inward migration expected for companions formed via disk fragmentation around massive stars.}
    \label{fig:fragment_trajectory}
\end{figure}

We analyze the fragment trajectory at the precise moment when the central temperature crosses the critical 2000~K threshold, representing the onset of SC formation. This time instance is shown in Figure~\ref{fig:fragment_trajectory}. We assume the SC has assembled approximately 5\% of the total fragment mass (corresponding to $M_{\rm sc}\simeq3.6\times10^{-2}\msun$), consistent with theoretical predictions and numerical simulations of low-mass star formation \citep{2013Tomida, 2013Vaytet}. While this mass fraction may seem modest, it represents the gravitationally bound core that will continue accreting material and eventually become a low-mass stellar companion. The exact mass ratio is less critical for the orbital dynamics since the central massive star ($M_*\approx 4.8\msun$) dominates the gravitational potential.

The color-coded curve in Figure~\ref{fig:fragment_trajectory} presents the computed orbital path of SC, where the progression from cold to hot colors represents the temporal evolution of the orbit over the 10-kyr integration period. The figure clearly illustrates both the highly eccentric nature of the orbit and the gradual inward spiraling motion. The red contour delineates the 3D fragment boundary based on our temperature-density criteria (see Appendix~\ref{sec:fragment_contour}), while the blue dashed circle indicates the Hill radius. The tiny white dashed circle marks the 5~AU averaging region used for velocity determination, and the magenta point identifies the location of maximum density within the fragment.

To determine the initial velocity conditions for trajectory integration, we employ a momentum-weighted averaging approach within a 5~AU sphere centered on the fragment's density maximum (black dashed circle). This method calculates the center-of-mass velocity by computing the total momentum of all gas elements within the averaging region and dividing by the total mass.

The orbital evolution is computed using fourth-order Runge-Kutta integration over 10 kyr with 0.01-year timesteps. The trajectory calculation incorporates the gravitational influence of the central star, whose mass increases according to the accretion history derived from the \citetalias{2020OlivaKuiper} simulation; because the stellar mass evolution is only available for about 10 kyr after the SC formation, our integration is likewise limited to that interval. 

Our analysis reveals a highly eccentric orbit with a periastron distance of approximately 17 AU and a late-stage eccentricity of $e\approx0.55$, indicating a significantly elliptical trajectory. The orbital evolution is inherently dynamic due to the continuously increasing stellar mass, which causes systematic changes in the orbital parameters over time. Most notably, we find a periastron decline rate of $-4.94 \times 10^{-3}$~AU~yr$^{-1}$ and we estimate that the companion will reach the stellar surface ($\approx$0.1~AU) in roughly $1.35\times10^4$~years. These timescales are significantly shorter than typical main-sequence lifetimes, suggesting that close companions formed through disk fragmentation may experience rapid inward migration. This has important implications for the formation of close binary systems and potentially the creation of contact binaries or stellar mergers around massive stars.

Near the end of our 10 kyr integration the instantaneous orbital period of the SC is on the order of $\sim$50~yr. Periodic or quasi-periodic variability on comparable (year-to-decade) timescales has been reported in some massive and intermediate-mass young objects in long-term infrared and maser monitoring campaigns. 
For example, the HMYSO G24.33+0.14 exhibits recurrent accretion-related flares with a well-established cycle of $\sim$8–9~yr, seen consistently in Class~II methanol masers and infrared light curves \citep{2019Wolak, 2022Hirota, 2023Liu}. Similarly, M17~MIR shows two, $\sim$decade-long accretion outbursts with quiescent interval of about six years \citep{2021Chen}. Moreover, a possible earlier near-IR brightening of S255IR–NIRS3 based on sparse archival photometry is suggested by \citet{2023Fedriani}, but that interpretation is debated given the absence of photometric error margins and corroborating high-resolution ejecta signatures.
For such systems, bursts separated by more than a decade may observationally appear as distinct, individual events. 
A bound secondary with a $\sim$decadal orbit can naturally produce such signatures by periodically modulating the inner-disk accretion flow or the irradiation geometry. We therefore note that the orbital periods implied by our trajectory — and their evolution as the companion migrates inward — may be directly relevant to interpreting long-term photometric or maser periodicities.

\subsection{Method Limitations and Uncertainties}

Although the hydro simulation includes the self-gravity of the disk, our trajectory analysis omits the gravitational field of the surrounding disk, which introduces an important source of error. In massive, self-gravitating disks, which can contribute 10–50\% of the total system mass \citep[e.g.,][]{2011Kuiper, 2016Kratter, 2018MeyerKuiper}, non-axisymmetric structures (spiral arms, global modes, and strong overdensities) produce large, time-dependent torques that can accelerate inward migration (and in some regimes drive rapid, Type-III or runaway migration), whereas in other regimes corotation torques or gap formation can slow migration and tie the object to the viscous evolution of the disk. In short, including the disk gravity can either increase or decrease the periastron decay rate depending on the fragment mass, local surface density, disk aspect ratio, and whether the fragment opens a gap or resides in a massive co-orbital region. Because our point-mass integration neglects these effects, the sign and magnitude of the error in the periastron decline are uncertain, and our orbital decay estimates should be regarded as provisional.

Additionally, the analysis does not consider gravitational interactions between multiple fragments, even though such fragments often form together in unstable disks \citep[e.g.,][]{2018VorobyovElbakyan}. Encounters between fragments can lead to binary formation via three-body interactions, orbital scattering that modifies trajectories, or mergers that change the mass distribution. These processes may either accelerate infall through dynamical heating or help stabilize wide orbits by redistributing angular momentum.

Hydrodynamic effects are also neglected, including gas drag and feedback between the fragment and the surrounding disk. Type I and II orbital migration, driven by gravitational torques with spiral density waves, can occur on timescales of $10^3-10^4$ years, potentially shorter than the integration period used here. Ram pressure from the fragment’s motion through the disk could further influence its internal structure and accretion rate.

The assumption of a fixed 5\% mass fraction for the companion does not reflect the reality of ongoing accretion. Both observations and theory suggest accretion rates of $10^{-4}-10^{-3}\msunyr$ \citep{2013Beuther, 2017Beuther, 2021Moscadelli, 2024WellsBeuther, 2024ZhangCyganowski, 2025SchneiderBeuther}, which could double the companion’s mass over $10^3-10^4$ years. Such mass growth would affect the gravitational parameter in the orbital calculations and could help stabilize wider orbits against inward migration.

Finally, the analysis omits the effects of stellar winds, radiation pressure, and magnetic fields, all of which become increasingly important as the central star evolves toward the main sequence. These feedback processes can unbind weakly bound companions and significantly reshape the disk, further altering the gravitational environment experienced by the fragment. Despite these limitations, our simplified approach provides valuable first-order estimates for companion formation processes.

\section{Power-law fits to azimuthally averaged midplane profiles} \label{sec:power_law}
To quantify how resolving the inner few AU alters the disk structure, we fit power laws to the azimuthally averaged midplane density and temperature profiles in two radial ranges: the inner region (2–30 AU) — now resolved in the refined model but unresolved in OK20 — and the outer disk (30–500 AU), which corresponds to the domain primarily sampled by OK20 and where the global disk structure dominates. We exclude $r<2$ AU to avoid inner–boundary (sink) effects. Before the burst (top panel, $t=8590$ yr) the inner density profile is very steep ($\rho\propto r^{-2.65}$), indicating a strongly centrally concentrated inner region in the refined run, while the outer disk is much shallower ($\rho\propto r^{-0.63}$), consistent with a relatively flat, extended mass distribution. The corresponding temperature slopes are moderately negative: the inner region follows $T\propto r^{-0.67}$, and the outer disk $T\propto r^{-0.42}$, values that are broadly compatible with a hot, actively heated inner region (advective/shock heating and reprocessing) plus irradiation-dominated outer disk.

As the system approaches the burst (second panel in Fig.~\ref{fig:profiles} — here the refined run has already undergone the burst while OK20 has not), the inner density slope flattens substantially ($\rho\propto r^{-1.15}$ in the 2–30 AU range), signalling mass redistribution: material that had been concentrated near the star is being spread by tidal stripping and by the formation of a more extended inner structure. At the same time the outer slope remains modestly negative ($\rho\propto r^{-0.78}$), showing that the outer disk is less affected by this rapid inner restructuring. Temperature slopes become slightly shallower in the inner region ($T\propto r^{-0.44}$), consistent with stronger, more spatially extended heating during and immediately after the mass deposition phase.

Near the peak (third panel in Fig.~\ref{fig:profiles}) both density and temperature slopes moderate again: the inner density sits at $\sim r^{-1.12}$, and the outer density shows a steeper value of $\sim r^{-1.58}$, indicating that the burst has injected significant energy and re-shaped the radial mass gradient across a broad radial range. Temperature slopes near peak are still typical of strongly heated inner regions ($T\propto r^{-0.69}$ in the inner disk) and shallower in the outer disk ($T\propto r^{-0.42}$), reflecting intense inner dissipation and outward radiative transport.

About 40 years after the burst (bottom panel in Fig.~\ref{fig:profiles}) the inner density steepens again ($\rho\propto r^{-2.02}$ in the 2–30 AU) while the outer disk relaxes to a shallower profile ($\rho\propto r^{-0.93}$). This re-steepening of the inner density is most plausibly interpreted as the inner region re-concentrating mass into a compact disk after the transient redistribution during disruption, while the outer disk cools and relaxes. Temperature slopes after 40 yr are comparatively flat (around $-0.39$ to $-0.43$) across both ranges, consistent with the disk settling to a state where irradiation and residual local dissipation set a moderate radial gradient.

\end{appendix}

\end{document}